\documentclass[journal]{IEEEtran}

\usepackage{booktabs}
\usepackage{xcolor}
\usepackage{colortbl}
\usepackage{amssymb}
\usepackage{algorithmic}
\usepackage{amsmath} 
\usepackage{amsfonts}
\usepackage[linesnumbered,ruled,lined]{algorithm2e}
\ifCLASSINFOpdf
\usepackage[pdftex]{graphicx}
\else
\usepackage[dvips]{graphicx}
\fi
\usepackage{epstopdf}
\usepackage{array}
\ifCLASSOPTIONcompsoc
\usepackage[caption=false,font=normalsize,labelfont=sf,textfont=sf]{subfig}
\else
\usepackage[caption=false,font=footnotesize]{subfig}
\fi
\usepackage{float} 
\usepackage{fixltx2e}
\usepackage{color}
\usepackage{xcolor}
\usepackage{textcomp}
\usepackage[normalem]{ulem}
\usepackage{cite}
\usepackage{multibib} % necessary for BSTcontrol
\usepackage{nomencl}
\makenomenclature
\usepackage{etoolbox}
\renewcommand\nomgroup[1]{%
	\item[\bfseries
	\ifstrequal{#1}{A}{Indices and Sets}{%
		\ifstrequal{#1}{B}{Continuous Decision Variables}{%
			\ifstrequal{#1}{C}{Discrete Decision Variables}{       \ifstrequal{#1}{D}{Parameters}{
			        \ifstrequal{#1}{E}{Dynamic System Variables and Parameters}} }}} ]}
\usepackage{hyperref}
\hypersetup{
	colorlinks=true,
	linkcolor=blue,
	filecolor=magenta,      
	urlcolor=cyan,}
\IEEEoverridecommandlockouts

\begin{document}
\bstctlcite{BSTcontrol}%externally control some aspects of the bibliography style (use multibib)

\title{Encoding Frequency Constraints in Preventive Unit Commitment Using Deep Learning with Region-of-Interest Active Sampling}

\author{Yichen~Zhang,~\IEEEmembership{Senior Member,~IEEE,}
	~Hantao~Cui,~\IEEEmembership{Senior Member,~IEEE,}
	~Jianzhe~Liu,~\IEEEmembership{Member,~IEEE,}
	~Feng~Qiu,~\IEEEmembership{Senior Member,~IEEE,}
	~Tianqi~Hong,~\IEEEmembership{Member,~IEEE,}
	~Rui~Yao,~\IEEEmembership{Senior Member,~IEEE}
 	~Fangxing~(Fran)~Li,~\IEEEmembership{Fellow,~IEEE}
	\thanks{
		This work was supported by the U.S. Department of Energy Office of Electricity -- Advanced Grid Modeling Program.
		
		Y. Zhang, J. Liu, F. Qiu, T. Hong, R. Yao are with Energy System Division, Argonne National Laboratory, Lemont, IL 60439 USA. (Email: yichen.zhang@anl.gov, jianzhe.liu@anl.gov, fqiu@anl.gov, thong@anl.gov, ryao@anl.gov). 
		
	    H. Cui is with  School of Electrical and Computer Engineering, Oklahoma State University, Stillwater, 74078 OK, USA. (Email: h.cui@okstate.edu)
		
	    F. Li is with Department of EECS University of Tennessee, Knoxville TN 37909 USA. (Email: fli6@utk.edu)
}}

\markboth{This paper has been accepted by IEEE TRANSACTIONS ON POWER SYSTEMS (DOI: 10.1109/TPWRS.2021.3110881)}%
{Shell \MakeLowercase{\textit{et al.}}: Bare Demo of IEEEtran.cls for IEEE Journals}
\maketitle

\begin{abstract}
With the increasing penetration of renewable energy, frequency response and its security are of significant concerns for reliable power system operations. Frequency-constrained unit commitment (FCUC) is proposed to address this challenge. Despite existing efforts in modeling frequency characteristics in unit commitment (UC), current strategies can only handle oversimplified low-order frequency response models and do not consider wide-range operating conditions. This paper presents a generic data-driven framework for FCUC under high renewable penetration. Deep neural networks (DNNs) are trained to predict the frequency response using real data or high-fidelity simulation data. Next, the DNN is reformulated as a set of mixed-integer linear constraints to be incorporated into the ordinary UC formulation. In the data generation phase, all possible power injections are considered, and a region-of-interest active sampling is proposed to include power injection samples with frequency nadirs closer to the UFLC threshold, which enhances the accuracy of frequency constraints in FCUC. The proposed FCUC is investigated on the IEEE 39-bus system. Then, a full-order dynamic model simulation using PSS/E verifies the effectiveness of FCUC in frequency-secure generator commitments.
\end{abstract}

% \fi
%\IEEEpeerreviewmaketitle

% symbol lists
\mbox{}
\nomenclature[A,01]{$i$,$j$, $\mathcal{N}_{\text{B}}$, $N_{\text{B}}$}{index, set, and number of buses}
\nomenclature[A,02]{$l$, $\mathcal{N}_{\text{L}}$, $N_{\text{L}}$}{index, set, and number of lines}
\nomenclature[A,03]{$k$, $\mathcal{N}_{\text{D}}$, $N_{\text{D}}$}{index, set, and number of loads}
\nomenclature[A,04]{$g$, $\mathcal{N}_{\text{G}}$, $N_{\text{G}}$}{index, set, number of synchronous generators}
\nomenclature[A,05]{$w$, $\mathcal{N}_{\text{W}}$, $N_{\text{W}}$}{index, set, number of wind turbine generators}
\nomenclature[A,06]{$m$, $\mathcal{N}_{\text{Y}}$, $N_{\text{Y}}$}{index, set, number of hidden layers of a neural network}
\nomenclature[A,07]{$n$, $\mathcal{N}_{\text{O}}$, $N_{\text{O}}$}{index, set, number of neurons in a layer}
\nomenclature[A,08]{$s$, $\mathcal{N}_{\text{S}}$, $N_{\text{S}}$}{index, set, number of samples}
\nomenclature[A,09]{$t$, $\mathcal{N}_{\text{T}}$, $N_{\text{T}}$}{index, set, number of period}
\nomenclature[A,10]{$\sigma_{b}(i)$}{index set of all buses connected to bus $i$}
\nomenclature[A,11]{$\sigma_{g}(i)$}{index set of all generators connected to bus $i$}
\nomenclature[A,12]{$\sigma_{w}(i)$}{index set of all wind turbines connected to bus $i$}
\nomenclature[A,13]{$\sigma_{d}(i)$}{index set of all loads connected to bus $i$}

\nomenclature[B,01]{$P_{g,t}^{\text{G}}$,$Q_{g,t}^{\text{G}}$}{power output of generator $g$ during period $t$}
\nomenclature[B,02]{$p_{g,t}^{\text{G}}$}{incremental output of generator $g$ from its minimum during period $t$}
\nomenclature[B,03]{$R_{g,t}^{\text{G}}$}{reserve of generator $g$ during period $t$}
\nomenclature[B,04]{$P_{l,t}$, $Q_{l,t}$}{active, reactive power flow on line $l$ during period $t$}
\nomenclature[B,06]{$V_{i,t}$}{voltage of bus $i$ during period $t$}

\nomenclature[C,01]{$u_{g,t}^{\text{G}}$}{1 if unit $g$ is scheduled on during period $t$ and 0 otherwise}

\nomenclature[D,01]{$P_{k,t}^{\text{D}}$, $Q_{k,t}^{\text{D}}$}{active, reactive power of load $k$ during period $t$}
\nomenclature[D,02]{$P_{w,t}^{\text{W}}$}{power output of wind turbine generator $w$ during period $t$}
\nomenclature[D,03]{$\underline{P}_{l}$, $\overline{P}_{l}$}{min, max active power flow of line $l$}
\nomenclature[D,04]{$\underline{Q}_{l}$, $\overline{Q}_{l}$}{min, max reactive power flow of line $l$}
\nomenclature[D,05]{$\underline{P}^{\text{G}}_{g}$, $\overline{P}^{\text{G}}_{g}$}{min, max active power output of unit $g$}
\nomenclature[D,06]{$\underline{Q}^{\text{G}}_{g}$, $\overline{Q}^{\text{G}}_{g}$}{min, max reactive power output of unit $g$}
\nomenclature[D,06]{$\underline{V}$, $\overline{V}$}{min, max allowable voltages}
\nomenclature[D,07]{$\bold{W}_{m}$, $\bold{b}_{m}$}{weight and bias of layer $m$ in a neural network}
\nomenclature[D,08]{$\lambda^{\text{F}}_{g}$}{fixed cost of unit $g$ at the point of $\underline{P}^{\text{G}}_{g}$}
\nomenclature[D,09]{$\lambda^{\text{M}}_{g}$,$\lambda^{\text{S}}_{g}$}{marginal, start-up cost of unit $g$}
\nomenclature[D,10]{$G_{ij}$, $B_{ij}$}{conductance and susceptance of line connected by bus $i$ and $j$}
\printnomenclature[0.7in]

\section{Introduction}\label{sec_intro}
%Frequency security is concerned at three progressive levels, that is, adequate security, steady-state security and dynamic security. 
%The \emph{adequate security}, which is known as frequency stabilization, requires that adequate spinning reserves should be scheduled all the time to regain power balance once a disturbance occurs such that the frequency can reach to a steady state \cite{Restrepo2005}. 
%The \emph{steady-state security} seeks to contain the steady-state frequency in the so-called continuous operation zone, which can be achieved primarily by droop gain scheduling \cite{Dvorkin2018}. 
%The \emph{dynamic security} probes into strategies to ensure that the frequency deviation should stay within permissible ranges to avoid unnecessary\footnote{The unnecessary relay actions refers to the scenarios where the system has adequate capacity to reach a viable steady state but the relays are triggered due to larger transients in frequency response.} frequency relay actions \cite{Zhang2019}. This is the most challenging requirement yet becomes increasingly important, and is the focus of this paper. 

%Traditionally, dynamic frequency security can be verified using simulation given a scheduling result. If insecure responses are found, the scheduling result will be refined. Such a procedure is flexible to accommodate different models \cite{Gu2018,Ahmadyar2018}, but relies on heuristic refinement algorithms, resulting in suboptimal solutions. 

With the increasing penetration of renewable energy, power system frequency stability is of significant concern for reliable system operations.  
Federal Energy Regulatory Commission (FERC) has suggested that conventional unit scheduling, commitment and dispatch will need to take into account primary and secondary frequency control capabilities in addition to the traditional economic and security constraints \cite{LBNL_frequency_metric_operation_planning}.
The frequency-constrained unit commitment (FCUC) is proposed to address this challenge \cite{Restrepo2005}. There are mainly two types of frequency nadir approximation models. The first approach uses the swing equation with a piecewise linear mechanical power approximation. In this case, the mechanical power is not governed by a closed-loop control but is assumed to follow a piecewise linear function with respect to time. Then, different versions of frequency security conditions can be derived and embedded into the optimization formulation. With this model, the frequency constrained stochastic economic dispatch is considered in \cite{Lee2013}. The optimal power flow with primary frequency response adequacy constraint is investigated in \cite{Chavez2014}, where the minimum required governor response is derived. The stochastic unit commitment (UC) is studied in \cite{Teng2015}. Ref. \cite{Prakash2018} employed the formulation in \cite{Teng2015} for the interval scheduling problem. In \cite{Wen2016}\cite{Badesa2019}\cite{Chu2020}, multiple frequency services are considered, where the supported power is approximated by integrator dynamics. However, such a model does not capture the interaction between frequency and mechanical power dynamics. The second type employs the so-called system frequency response (SFR) model. Refs. \cite{Anderson1990}, \cite{Ahmadi2014} derived the analytical expression of the SFR under a step input and proposed an efficient piecewise linearization that converts the analytical solution into a set of constraints. This strategy has been fully extended into microgrid scheduling \cite{Gholami2018}, hierarchical frequency control \cite{Muzhikyan2018}, optimal power flow \cite{Nguyen2019}, and unit commitment with renewable energy resources \cite{Zhang2020b}, respectively.
Despite these excellent efforts of modeling SFR in mathematical programming-based (MP-based) scheduling, these methods rely on the low-order model approximation that cannot be able to capture the entire SFR characteristics. In the classic SFR \cite{Anderson1990}\cite{Ahmadi2014}, individual generator speeds deviate from the center of inertia (COI) frequency depending on the electric distances of generators. Therefore, only considering COI frequency can ignore insecure responses of certain generators. These methods cannot incorporate high-order models since there are no analytical solutions for the step response. In addition, nonlinearities in SFR such as deadbands and saturations cannot be taken into considerations by the existing approaches, which, however, have significant impacts on SFR \cite{Kou2016}. On the other hand, directly incorporating frequency dynamics will result in highly complex MP problems \cite{Avramidis2020}\cite{Zhao2020c}.

A pioneering data driven approaches have been proposed in \cite{Lagos2021} for FCUC, where a classification decision tree for frequency security was designed. In this paper, we propose a deep neural network (DNN)-based trajectory constraint encoding framework as deep neural networks (DNN) have shown strong representation power to amend the limitations of model-based approaches \cite{Gupta2019,Sun2018a,Du2020a}. 
The framework will first train a DNN-based frequency nadir predictor using real operation data or high-fidelity simulation data. In this case, the frequency nadir predictor can reflect a variety of model types and corresponding nonlinearities. Then, the DNN will be reformulated into a set of mixed-integer linear constraints and further incorporated into the unit commitment program. It is worth mentioning that this re-formulation is exact if the rectified linear unit (ReLU) is employed as the activation function \cite{Anderson2019}. The idea has been proposed in our previous work in \cite{Zhang2021}. But it is worth mentioning that the FCUC that we are addressing is a significantly different problem compared with the previous work in \cite{Zhang2021} in terms of transmission system's stability, heterogeneity, and sampling complexity under varying operating conditions. The methodology and working pipeline in \cite{Zhang2021} cannot be directly applied. A systematic methodology and working pipeline are needed. The challenges of the FCUS and our contribution are concluded as follows 
\begin{enumerate} 
	\item First, although both papers consider post-contingency system performance, \cite{Zhang2021} does not consider the system stability issue. Post-contingency stability issue includes transient stability, small-signal stability, and voltage stability, and they include non-trivial difficulty to scheduling problems as directly training the frequency nadir predictor over all scenarios is not applicable. To tackle this issue, we build a series of post-processing criteria to group different scenarios. Then, we use another neural network to predict the stability and filter out unstable operating conditions for the frequency nadir predictor.
	\item Secondly, unlike \cite{Zhang2021}, we consider varying disturbance locations and heterogeneous responses from generators. Feature construction to take the heterogeneity into account is essential for the prediction performance. In this paper, based on the generator frequency response characteristics and empirical studies, we construct appropriate and concise feature vectors and propose associated embedding mixed-integer linear program (MILP) formulations.
	\item Thirdly, in this paper, we propose a new region-of-interest active sampling method to allow us to consider wide-range operation conditions to ensure the reliability of our method. This active sampling method allows us to select the most informative samples without labeling efforts and thus resolves the labeling bottleneck.
\end{enumerate}

The limitations of existing methods and the improvements of our approach can be concluded as follows
\begin{itemize}
	\item As discussed before, existing methods are not sufficient in capturing the frequency dynamics and not capable of embedding more complicated models. Our data-driven approach overcomes this challenge.
	\item Existing methods will introduce considerable integer variables since they rely on piecewise linearization of the analytical frequency nadir for constraint embedding. Additionally, their embedding procedure will introduce approximation error. Although our approach will introduce integer variables, there is no approximation error during our embedding procedure since the reformulation of the DNN is exact if the ReLU is used.
	\item Compared with our method, existing model-based methods cannot consider possible unstable scenarios.
\end{itemize}

The remainder of the paper is organized as follows. 
Section \ref{sec_prob} describes the problem setup. 
Section \ref{sec_dnn} describes the DNN-based predictors for frequency trajectory and stability with feature selections. 
Section \ref{sec_al_sampling} describes the data generation using the region-of-interest active sampling strategy.
Section \ref{sec_uc} express the FCUC formulation, followed by the case study in Section \ref{sec_case}, discussions in Section \ref{sec_discuss}, and conclusions in Section \ref{sec_con}.

\section{Problem Statement}\label{sec_prob}
Given the load forecast $\bold{d}$, the ordinary UC problem in Eq. (\ref{eq_uc}) aims to calculate the optimal dispatch command $\bold{u}\in\mathcal{U}$ such that the operating cost is minimized while the system states $\bold{s}$ respect the constraints
\begin{align}
\label{eq_uc}
\begin{aligned}
&\min_{\bold{u}_{t}\in\mathcal{U}}\quad C(\bold{s}_{t},\bold{u}_{t})\\
&\text{s.t.}\quad \bold{f}_{\text{u}}(\bold{s}_{t},\bold{u}_{t},\bold{d}_{t})=0,\bold{g}_{\text{u}}(\bold{s}_{t},\bold{u}_{t},\bold{d}_{t})\leq 0\quad\forall t\in\mathcal{T}
\end{aligned}
\end{align}
where $\bold{f}_{u}$ and $\bold{g}_{u}$ denotes the equality and inequality constraints, respectively.  Assume a disturbance $\varpi$ occurs at scheduling period $\tau$, which results in power imbalance and the decline of frequency. Let the frequency nadir of the system be denoted as $f_{\text{ndr}}$, which can be assumed to admit a nonlinear relation with respect to the disturbance, system states, dispatch command, and load condition
\begin{align}
\label{eq_nadir_exp}
f_{\text{ndr}} = h^{\text{f}}(\bold{s}_{\tau},\bold{u}_{\tau},\bold{d}_{\tau},\varpi)
\end{align}
The frequency-constrained UC (FCUC) searches the optimal dispatch command such that the resulting system states ensure a secure frequency response for any disturbance from a pre-defined set $\mathcal{W}$ occurring within the scheduling period $\mathcal{T}$. The FCUC reads as follows
\begin{subequations}
\label{eq_fcuc}
\begin{align}
&\min_{\bold{u}_{t}\in\mathcal{U}}\quad C(\bold{s}_{t},\bold{u}_{t})\\
&\text{s.t.}\quad \bold{f}_{\text{u}}(\bold{s}_{t},\bold{u}_{t},\bold{d}_{t})=0,\bold{g}_{\text{u}}(\bold{s}_{t},\bold{u}_{t},\bold{d}_{t})\leq 0\quad\forall t\in\mathcal{T}\\ 
&\quad\quad h^{\text{f}}(\bold{s}_{t},\bold{u}_{t},\bold{d}_{t},\varpi)\geq f_{\min}\quad\forall t\in\mathcal{T},\forall \varpi\in\mathcal{W}\label{eq_fcuc_3}
\end{align}
\end{subequations}
where $\bold{f}_{u}$ and $\bold{g}_{u}$ denotes the equality and inequality constraints, respectively. The disturbance set $\mathcal{W}$ is defined to be the $N-1$ generator loss. It is almost guaranteed that the largest generator outage will result in the worse frequency nadir. Thus, we consider the worst-case contingency in the FCUC problem to be the loss of the generator with the largest power output, which changes within the scheduling period $\mathcal{T}$.

\begin{figure}[h]
	\centering
	\includegraphics[scale=0.09]{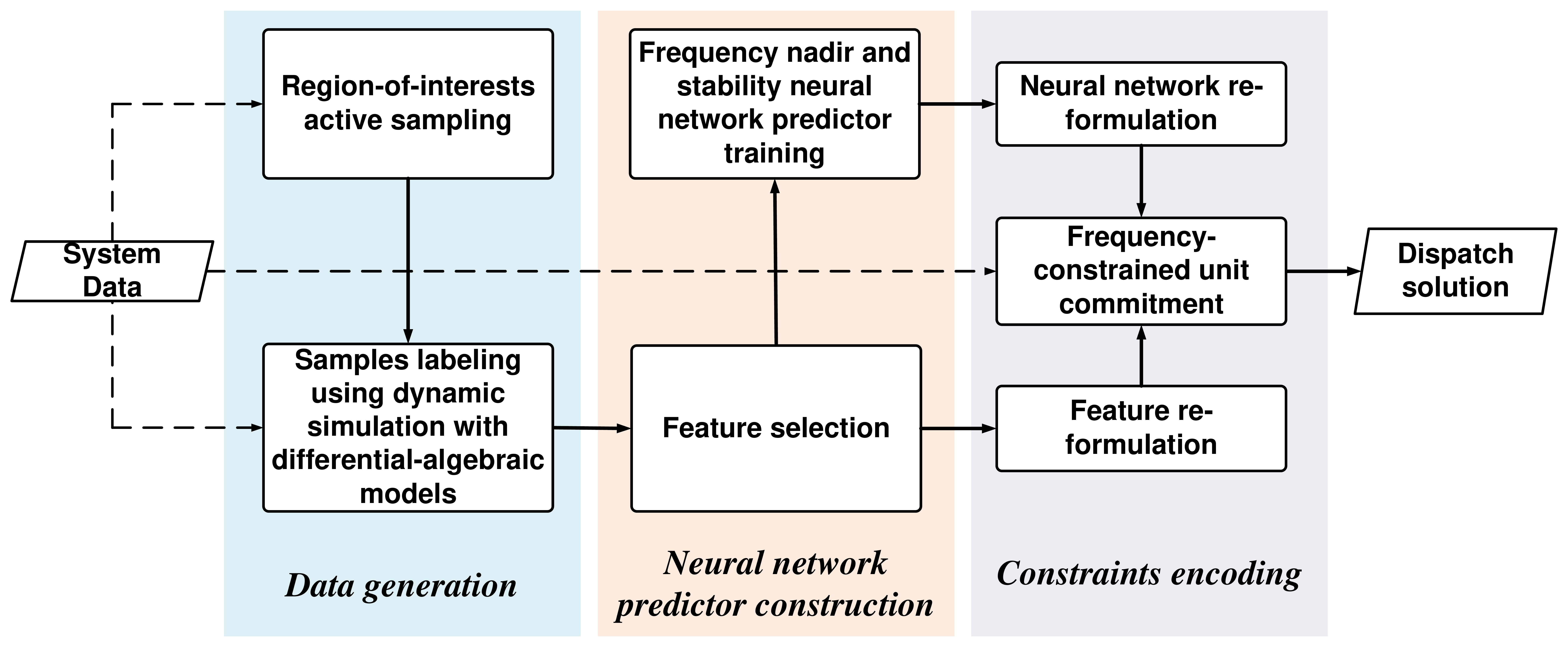}
	\vspace{-0.15in}
	\caption{Overview of the proposed FCUC framework and working pipeline.}
	\label{fig_FCUC_Strategy}
\end{figure}

The underlying assumption for the existence of (\ref{eq_nadir_exp}) is that the system after the worst-case disturbance will admit a stable trajectory. Since we are considering the full power injection space, this assumption may not hold. There are two prerequisites for us to consider the frequency trajectory constraint: (1) the post-disturbance system should have a power flow solution, which is closely related to voltage stability; (2) the post-disturbance system should hold the rotor angle stability condition. In other words, Eq. (\ref{eq_fcuc_3}) may provide incorrect outputs under voltage and rotor angle instable cases and mislead the feasibility of the entire UC program in (\ref{eq_fcuc}). To tackle this issue, we assume there exist a function that can map from $\bold{s}_{t},\bold{u}_{t},\bold{d}_{t}$ to the probability of stability under the worst-case disturbance as shown in (\ref{eq_fcuc_stab})
\begin{subequations}
\label{eq_fcuc_stab}
\begin{align}
&\min_{\bold{u}_{t}\in\mathcal{U}}\quad C(\bold{s}_{t},\bold{u}_{t})\\
&\text{s.t.}\quad \bold{f}_{\text{u}}(\bold{s}_{t},\bold{u}_{t},\bold{d}_{t})=0,\bold{g}_{\text{u}}(\bold{s}_{t},\bold{u}_{t},\bold{d}_{t})\leq 0\\ 
&\quad\quad h^{\text{f}}(\bold{s}_{t},\bold{u}_{t},\bold{d}_{t},\varpi)\geq f_{\min}\label{eq_fcuc_stab_3}\\
&\quad\quad h^{\text{p}}(\bold{s}_{t},\bold{u}_{t},\bold{d}_{t},\varpi)_{[2]}\geq h^{\text{p}}(\bold{p}_{t},\bold{u}_{t},\bold{d}_{t},\varpi)_{[1]}\label{eq_fcuc_stab_4}\\
&\quad\quad\forall t\in\mathcal{T}, \varpi\in\mathcal{W}
\end{align}
\end{subequations}
where $h^{\text{p}}$ is the probabilistic stability descriptor that outputs a two-dimensional real-valued vector. The first entry denotes the predicted probability of a case being unstable, and the second entry denotes the predicted probability of a case being stable. The subscript $[s]$ denotes the $s$th entry of the output vector. The ultimate goal of this paper is to construct (\ref{eq_fcuc_stab}), which will be described in the following sections using the working pipeline shown in Fig. \ref{fig_FCUC_Strategy}.

\section{Deep Learning-Based Trajectory Constraint Approximation}\label{sec_dnn}

\subsection{DNN-Based Frequency Nadir Predictor}\label{sec_dnn_sub_freq}
Let this neural network frequency nadir predictor for $h^{\text{f}}$ be expressed as
\begin{align}
\label{eq_dnn}
\hat{f}_{\text{ndr}} = \hat{h}^{\text{f}}(\bold{x}^{\text{f}};\bold{W}^{\text{f}},\bold{b}^{\text{f}})
\end{align}
where $\bold{x}^{\text{f}}$ denotes the feature vector, and $\bold{W}^{\text{f}}$ and $\bold{b}^{\text{f}}$ denote the neural network parameters to be trained. To fully represent all possible operating conditions, a sufficient sample set $\mathcal{N}_{\text{S}}$ will be created\footnote{The subscripts $s$ and $t$ are essentially the same as they both denote different system operating scenarios. Following the convention, we use $s$ in the machine learning problem and $t$ in the UC problem.}. Let $f_{\text{ndr},s}$ and $\hat{f}_{\text{ndr},s}$ denote the actual and predicted frequency nadirs of sample $\bold{x}^{\text{f}}_{s}$.
%Let $\bold{X}^{\text{f}}$ denote the input samples to the DNN shown in (\ref{eq_dnn}): Each row represents a data sample, and each column represents different samples of one variable, shown as follows
%\begin{align}
%\label{eq_dnn_x}
%\bold{X}^{\text{f}} &= \left[\begin{array}{c}(\bold{x}^{\text{f}}_{1})^{T},(\bold{x}^{\text{f}}_{2})^{T} ,\cdots, (\bold{x}^{\text{f}}_{s})^{T}, \cdots, (\bold{x}^{\text{f}}_{N_{\text{S}}})^{T} \end{array}\right]^{T}
%\end{align}
%Let a vector $\bold{y}$ denote the label output data, which consists of the frequency nadir of different samples expressed below
%\begin{align}
%\label{eq_dnn_y}
%\bold{y}^{\text{f}} = \left[\begin{array}{cccccc}
%f_{\text{ndr},1}^{*} & f_{\text{ndr},2}^{*} &\cdots & f_{\text{ndr},s}^{*}&\cdots&f_{\text{ndr},N_{\text{S}}}^{*}\\
%\end{array}\right]^{T}
%\end{align}
Now consider a fully connected neural network with $N^{\text{f}}_{\text{Y}}$ hidden layers. Each layer uses a rectified linear unit (ReLU) activation function denoted as $\sigma(\cdot)=\max(\cdot,0)$, and the output layer uses a linear activation function. The predicted nadir can be expressed as follows
\begin{subequations}
\label{eq_nn_freq}
\begin{align}
&\bold{z}_{1}^{\text{f}}=\bold{x}^{\text{f}}_{s}\bold{W}^{\text{f}}_{1}+\bold{b}^{\text{f}}_{1}\label{eq_nn_layer_in}\\
&\bold{\hat{z}}^{\text{f}}_{m}=\bold{z}^{\text{f}}_{m-1}\bold{W}^{\text{f}}_{m}+\bold{b}^{\text{f}}_{m}\label{eq_nn_layer_hidden}\\
&\bold{z}^{\text{f}}_{m}=\max(\bold{\hat{z}}^{\text{f}}_{m},0)\label{eq_nn_layer_ReLU}\\
&f_{\text{ndr},s}=\bold{z}^{\text{f}}_{N^{\text{f}}_{\text{Y}}}\bold{W}^{\text{f}}_{N^{\text{f}}_{\text{Y}}+1}+\bold{b}^{\text{f}}_{ N^{\text{f}}_{\text{Y}}+1}\label{eq_nn_layer_out}
\end{align}
\end{subequations}
where matrix $\bold{W}^{\text{f}}_{m}$ and vector $\bold{b}^{\text{f}}_{m}$ for $m=1,\cdots,N^{\text{f}}_{\text{Y}}$ represent the set of weight and bias across all hidden layers, and $\bold{W}^{\text{f}}_{N^{\text{f}}_{\text{Y}}+1}$ and $\bold{b}^{\text{f}}_{N^{\text{f}}_{\text{Y}}+1}$ represent the set of weight and bias of the output layer. We would like to minimize the total mean squared error between the predicted output and the labeled outputs of all samples as follows
\begin{align}
\min_{\bold{W}^{\text{f}}_{m}, \bold{b}^{\text{f}}_{m}}\frac{1}{N^{\text{f}}_{\text{S}}}\sum_{s=1}^{N^{\text{f}}_{\text{S}}}(f_{\text{ndr},s}-\hat{f}_{\text{ndr},s})^{2}
\end{align}
There are two essential prerequisites: (1) making correct choice of input features among $\bold{s}$, $\bold{u}$, $\bold{d}$ and $\varpi$; (2) generating representative samples under different system operating conditions.

\subsubsection{Feature Selection}
The frequency response from a synchronous generator (SG) is less dependent on its output unless there is not enough headroom, which will be avoided in the UC problem. In addition, we assume the droop control of the turbine governor for all SGs will not change. Therefore, it is sufficient to select the status of each SG $u_{g,s}^{\text{G}}$ as the input features. Both the magnitude and location of the disturbance will have impacts on the frequency response. Based on our discussion regarding the disturbance in Section \ref{sec_prob}, the disturbance magnitude can be expressed as
\begin{align}
\label{eq_dis_1}
\begin{aligned}
P^{\varpi}_{s}=\max_{g\in\mathcal{N}_{\text{G}}}(P_{1,s}^{\text{G}},\cdots,P^{\text{G}}_{g,s},\cdots,P^{\text{G}}_{N_{\text{G}},s})
\end{aligned}
\end{align}
We use the bus index to represent the location information as
\begin{align}
\label{eq_dis_2}
\begin{aligned}
g^{\varpi}_{s}=\arg\max_{g\in\mathcal{N}_{\text{G}}}(P_{1,s}^{\text{G}},\cdots,P^{\text{G}}_{g,s},\cdots,P^{\text{G}}_{N_{\text{G}},s})
\end{aligned}
\end{align}
We encode both information into one vector as
\begin{align}
\label{eq_dis_3}
\begin{aligned}
\bold{x}^{\varpi}_{s}=\left[\underbrace{0,\cdots,0}_{g^{\varpi}_{s}-1 \text{ element}}, \underbrace{P^{\varpi}_{s}}_{g^{\varpi}_{s}\text{th element}},0,\cdots,0\right]
\end{aligned}
\end{align}
Then, the feature vector of sample $s$ reads as 
\begin{align}
\label{eq_feature_freq}
\begin{aligned}
\bold{x}^{\text{f}}_{ s}=\left[u^{\text{G}}_{1,s},\cdots,u^{\text{G}}_{N_{\text{G}},s},\bold{x}^{\varpi}_{s}\right]
\end{aligned}
\end{align}

\subsection{DNN-Based Stability Predictor}\label{sec_dnn_sub_stab}
Similar to the frequency nadir predictor, we construct a neural network $\hat{h}^{\text{p}}$ to approximate the probabilistic stability descriptor $h^{\text{p}}$. To do this, we denote the classes of unstable and stable as $C_{q}$ for $q=1,2$, respectively. The label data after one-hot encoding reads $\bold{c}^{*}_{s} =c^{*}_{s,1},c^{*}_{s,2}$, where $c^{*}_{s,q}=1$ indicates that sample $s$ belongs to class $q$. We then apply probabilistic smoothing approximations to the discrete label values. It is well known that when the targets are one-hot encoded and an appropriate loss function is used, an neural network directly estimates the posterior probability of class membership $C_{q}$ conditioned on the input variables $\bold{x}^{\text{p}}$, denoted by $p(C_{q}|\bold{x}^{\text{p}})$. Then, the stability predictor can be expressed as follows 
\begin{align}
\label{eq_dnn_stab}
\bold{p}_{\text{stab}} =
\left[\begin{array}{cc}
p(C_{1}|\bold{x}^{\text{p}}),p(C_{2}|\bold{x}^{\text{p}})
\end{array}\right] = \hat{h}^{\text{p}}(\bold{x}^{\text{p}};\bold{W}^{\text{p}},\bold{b}^{\text{p}})
\end{align}
where $\hat{h}^{\text{p}}$ reads as follows
\begin{subequations}
\label{eq_nn_stab}
\begin{align}
&\bold{z}^{\text{p}}_{1}=\bold{x}^{\text{p}}_{s}\bold{W}^{\text{p}}_{1}+\bold{b}^{\text{p}}_{1}\\
&\bold{\hat{z}}^{\text{p}}_{m}=\bold{z}^{\text{p}}_{m-1}\bold{W}^{\text{p}}_{m}+\bold{b}^{\text{p}}_{m}\\
&\bold{z}^{\text{p}}_{m}=\max(\bold{\hat{z}}^{\text{p}}_{m},0)\\
&\bold{p}_{\text{stab},s}=\bold{z}^{\text{p}}_{N^{\text{p}}_{\text{Y}}}\bold{W}^{\text{p}}_{N^{\text{p}}_{\text{Y}}+1}+\bold{b}^{\text{p}}_{N^{\text{p}}_{\text{Y}}+1}
\end{align}
\end{subequations}
where $m=1,\cdots,N^{\text{p}}_{\text{Y}}$ represent the indices for all hidden layers, and $N^{\text{p}}_{\text{Y}}+1$ is the output layer. The network parameters can be calculated using the maximum likelihood estimation. Therefore, we minimize the negative logarithm of the likelihood function, known as the cross-entropy loss, as follows
\begin{align}
L=-\frac{1}{N^{\text{p}}_{\text{s}}}\sum_{s=1}^{N^{\text{p}}_{\text{s}}}\sum_{q=1}^{2}c^{*}_{s,q}\log p_{s}(C_{q}|\bold{x}^{\text{p}})
\end{align}

\subsubsection{Feature Selection}
The disturbance information is also essential for the stability predictor. The same encoding method in (\ref{eq_dis_1})-(\ref{eq_dis_3}) is used. Considering that only online SGs will respond to disturbances, their commitment status is also employed. The voltage stability is highly related to the loading condition of the system. Since we relax the generator Var power limits, the active power will have a dominant impact. The transient stability margin depends on the power-angle characteristics, which are determined by the active power injection. Thus, the active power injection of all SGs will be encoded into the feature vector. And the overall feature reads
\begin{align}
\label{eq_feature_stab}
\begin{aligned}
\bold{x}^{\text{p}}_{s}=[u^{\text{G}}_{1,s},\cdots,u^{\text{G}}_{N_{\text{G}},s}, \bold{x}^{\varpi}_{s},P_{1,s}^{\text{G}},\cdots,P^{\text{G}}_{N_{\text{G}},s} ]
\end{aligned}
\end{align}

%\begin{align}
%\label{eq_dnn_X_stab}
%\bold{X}^{\text{p}}
%&=\left[\begin{array}{ccccccccccc}
%u^{\text{G}}_{1,1} & \cdots & u^{\text{G}}_{N_{\text{G}},1} & \bold{x}^{\varpi}_{1} &  P_{1,1}^{\text{G}} & \cdots & P^{\text{G}}_{N_{\text{G}},1} \\
%u^{\text{G}}_{1,2} & \cdots & u^{\text{G}}_{N_{\text{G}},2} & \bold{x}^{\varpi}_{2} &  P_{1,2}^{\text{G}} & \cdots & P^{\text{G}}_{N_{\text{G}},2} \\
%\vdots & \ddots & \vdots & \vdots & \vdots & \ddots & \vdots \\
%u^{\text{G}}_{1,N_{\text{S}}} & \cdots & u^{\text{G}}_{N_{\text{G}},N_{\text{S}}} & \bold{x}^{\varpi}_{N_{\text{S}}} & P_{1,N_{s}}^{\text{G}} &\cdots & P^{\text{G}}_{N_{\text{G}},N_{s}} \\
%\end{array}\right] 
%\end{align}

\section{Data Generation and Region-of-Interest Active Sampling}\label{sec_al_sampling}
Consider a power network with $N_{\text{G}}$ SGs, $N_{\text{W}}$ WTGs, and $N_{\text{D}}$ loads. Let $\mathcal{N}_{\text{G}}$, $\mathcal{N}_{\text{W}}$, and $\mathcal{N}_{\text{D}}$ denote the set of SGs, WTGs and loads, respectively. The full dimension injection samples are 
\begin{align}
\label{eq_full_injection}
\begin{aligned}
\Phi = &[
\bold{P}_{1}^{\text{W}},\cdots,\bold{P}^{\text{W}}_{N_{\text{W}}},
\bold{P}_{1}^{\text{G}},\cdots,\bold{P}^{\text{G}}_{N_{\text{G}}},
\bold{Q}_{1}^{\text{G}},\cdots,\bold{Q}^{\text{G}}_{N_{\text{G}}},\\
&\bold{P}_{1}^{\text{D}},\cdots,\bold{P}^{\text{D}}_{N_{\text{D}}},
\bold{Q}_{1}^{\text{G}},\cdots,\bold{Q}^{\text{G}}_{N_{\text{D}}}]
\end{aligned}
\end{align}
where $\bold{P}_{g}^{\text{G}}$ and $\bold{Q}_{g}^{\text{G}}$ are samples of active and reactive power injections for SG $g$; $\bold{P}_{w}^{\text{W}}$ are samples of active power injections for WTG $w$; $\bold{P}_{k}^{\text{D}}$ and $\bold{Q}_{k}^{\text{D}}$ are samples of load active and reactive power injections for load $k$. Each sample in $\Phi$ is denoted as $\phi_{s}=\Phi_{[s,:]}$\footnote{The subscript $[s,:]$ denotes the $s$th row of a matrix.}, and the corresponding nadir (label) is denoted as $f_{\text{ndr},s}$.

\begin{figure}[h]
	\centering
	\includegraphics[scale=0.25]{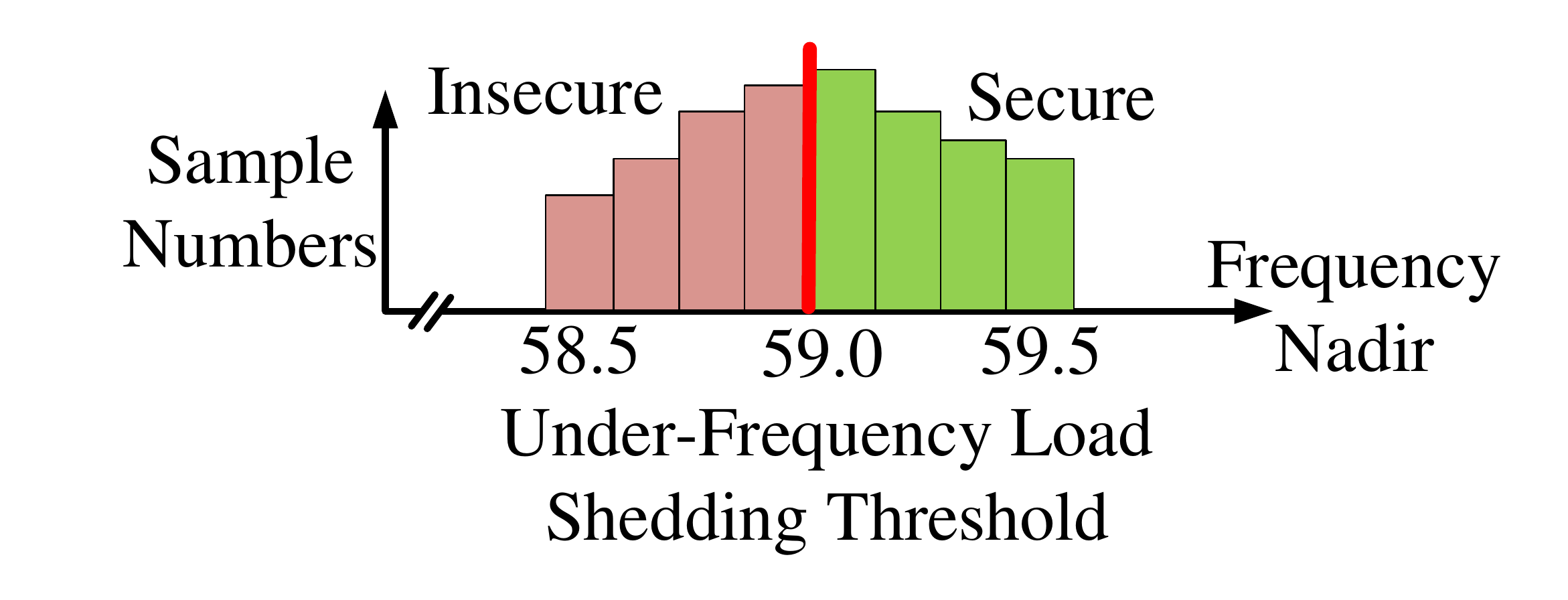}
	\vspace{-0.15in}
	\caption{A conceptual illustration of a desired training dataset.}
	\label{fig_Sample_Illustration}
\end{figure}

Unlike many works that sample around a narrow operation range, we consider the wide-range space of all power injections in (\ref{eq_full_injection}) to ensure reliability under vast ranges of operating conditions. In this case, not every sample admits a meaningful frequency nadir. In fact, many sampled power injections are not stable under the worst-case disturbance. Furthermore, what we are concerned about the most is the frequency nadir close to the underfrequnecy load shedding (UFLS) threshold. The rationale behind this is that a larger prediction error may not impact the UC result if the real frequency nadir is far from the UFLS, but a smaller error may cause inappropriate relay actions if the actual frequency nadir is close to the UFLS. When the actual frequency nadir is far from the UFLS threshold, whether the frequency security is a binding constraint or not to the FCUC will not be changed by the prediction error. Thus, under limited computation resources, it is reasonable to focus on improving the accuracy of frequency nadir prediction near the critical threshold. To train such a predictor, the distribution of the training dataset should contain denser samples where the frequency nadirs are closer to the UFLS threshold, as illustrated in Fig. \ref{fig_Sample_Illustration}. 
To generate such a training set, traditional methods will first perform uniform sampling over all power injections, label all samples to retrieve the nadir information, and then select labeled samples based on desired distributions. However, the labeling procedure needs time-domain simulation with full-order differential-algebraic equations. Labeling adequate samples to achieve desired distributions requires significant computation resources and time, which is known as the \emph{labeling bottleneck}. In other words, \emph{how to sample the power injections whose labels are within the region-of-interest without labeling efforts} is essential.

The proposed active sampling method can proceed as follows. We first define the sample as secure if the corresponding frequency nadir is above the UFLS threshold and insecure otherwise, as illustrated in Fig. \ref{fig_Sample_Illustration}. Second, we train a frequency security discriminator using the active learning method. The discriminator will output the predicted probability of a sample being secure. Then, we employ the active learning method to train this discriminator. Compared with the traditional passive learning approaches, the active learning method will actively select unlabeled samples, query their labels, and then perform the training on these data instances \cite{settles2009active}. During this procedure, samples whose label values are closer to the UFLS are easier to be selected. The rationale lies in the fact that it is more difficult for the discriminator to classify samples that are closer to the decision boundary, which is the UFLS threshold in our case. Therefore, we need to select samples whose posterior probability provided by the discriminator is the closest to 0.5 \cite{settles2009active}. In other words, the selected sample is the least confident to the discriminator.
\begin{algorithm}
	\SetKwData{Left}{left}\SetKwData{This}{this}\SetKwData{Up}{up}
	\SetKwFunction{Union}{Union}\SetKwFunction{Label}{Label}\SetKwFunction{Train}{Train}
	\SetKwInOut{Input}{input}\SetKwInOut{Output}{output}
	
	\Input{Labeled set $\mathcal{L}$, unlabeled set $\mathcal{U}$, sampling strategy $a(\cdot,\cdot)$, sampling batch size $B$, sample labeling function $\Label(\cdot)$, frequency security discriminator $\texttt{Clf}(\cdot)$, discriminator training function $\Train(\cdot,\cdot)$}
	
	%	\BlankLine
	$\mathcal{A} \leftarrow \emptyset$ \tcp*[f]{initialize the set to store acquisition samples}
	
	\Repeat{some stopping criterion}{
		
		$\texttt{Clf}(\cdot) \leftarrow \Train(\mathcal{L},\texttt{Clf}(\cdot))$ \tcp*[f]{train the discriminator using current $\mathcal{L}$}
		
		\For{$s \leftarrow 1$ \KwTo $B$}{
			
			$\phi_{s} \leftarrow \arg\max a(\mathcal{U}, \texttt{Clf}(\cdot))$ \tcp*[f]{active sampling using discriminator}
			
			$f_{\text{ndr},s},f_{\text{scr},s} \leftarrow \Label(\phi_{s})$
			
			$\mathcal{L} \leftarrow \mathcal{L}\cup\textless\phi_{s},f_{\text{scr},s}\textgreater$ 
			
			$\mathcal{A} \leftarrow \mathcal{A}\cup\textless\phi_{s},f_{\text{ndr},s}\textgreater$
			
			$\mathcal{U} \leftarrow \mathcal{U}-\phi_{s}$
		}
	}
	\Output{All acquisition instances $\mathcal{A}$}
	\caption{Region-of-Interest Active Sampling}\label{algo_roi_as}
\end{algorithm}
Let the frequency security discriminator be denoted as $\texttt{Clf}(\cdot)$, the UFLS threshold be denoted as $f_{\text{UFLS}}$, the security label of a sample $\phi_{s}$ be denoted as $f_{\text{scr}, s}$, where $f_{\text{scr}, s}=1$ if $f_{\text{ndr}, s}\geq f_{\text{UFLS}}$ and $f_{\text{scr}, s}=0$ otherwise. The details of the region-of-interest active sampling algorithm is illustrated in Algorithm \ref{algo_roi_as}. The overall strategy is introduced as follows
\begin{enumerate}
 	\item Define the frequency security discriminator $\texttt{Clf}(\cdot)$; Among all the samples $\Phi$, label a small portion denoted as $\mathcal{L}$ for frequency security discriminator training
 	\item Initialize the set denoted as $\mathcal{A}$ to store acquisition samples for frequency nadir predictor training $\hat{h}_{\text{f}}(\cdot)$
 	\item Train the discriminator using current labeled dataset $\mathcal{L}$
 	\item Perform the active sampling using the discriminator; The active sampling function $a(\mathcal{U}, \texttt{Clf}(\cdot))$ reads as follows
	 	\begin{align}
	 	\label{eq_entropy_sampling}
	 	\begin{aligned}
	 	\phi_{s}&=\underset{\mathcal{U}}{\mathrm{argmax}}\left( a(\mathcal{U}, \texttt{Clf}(\cdot))\right) \\
	 	&=\underset{\mathcal{U}}{\mathrm{argmax}}\left( -\sum_{i=0}^{1}p(i|\mathcal{U})\log p(i|\mathcal{U})  \right) 
	 	\end{aligned}
	 	\end{align}
	 	where $p(i|\mathcal{U})$ is the predicted posterior probability of class membership (0 for insecure and 1 for secure) conditioned on the sample inputs.
	 \item Retrieve both nadir and security labels; Add the nadir label $f_{\text{ndr},s}$ with the acquired sample $\phi_{s}$ to $\mathcal{A}$; Add the security label $f_{\text{scr},s}$ with the acquired sample $\phi_{s}$ to $\mathcal{L}$; Remove the sample from the unlabeled pool.
	 \item Repeat (4) and (5) to reach the batch number; Repeat (3)-(5) for the entire process
\end{enumerate}
It is worth mentioning that Eq. (\ref{eq_entropy_sampling}) entails only a simple sorting problem that finds the largest from a finite set of numerical values. Efficient algorithms to solve such problems have been extensively studied, especially in the realm of computer science. To select the most informative samples globally, we will perform this operation on all samples. But the algorithm can be flexible to operate on randomly grouped subsets of the overall sample set to increase the computation time if the entire sample size is too large.
The ultimate goal of Algorithm \ref{algo_roi_as} is to generate the dataset $\mathcal{A}$ for training the frequency nadir predictor $\hat{h}_{\text{f}}$. By calculating the entropy of the security discriminator outputs and selecting the extreme value, the corresponding samples will be closer to the decision boundary, that is, the UFLS threshold.

\section{Frequency Constrained Unit Commitment}\label{sec_uc}

In this section, we will introduce the FCUC formulation. Consider an $N_{\text{B}}$-bus power network. Consider the scheduling period from 1 to $N_{\text{T}}$. Each SG should be dispatched with the permissible limits
\begin{subequations}
	\label{eq_gen_dispatch}
	\begin{align}
	&0\leq p_{g,t}^{\text{G}} \leq (\overline{P}^{\text{G}}_{g} - \underline{P}^{\text{G}}_{g}) u_{g,t}^{\text{G}} \quad\forall g\\
	&\underline{Q}^{\text{G}}_{g} u_{g,t}^{\text{G}} \leq Q_{g,t}^{\text{G}} \leq \overline{Q}^{\text{G}}_{g} u_{g,t}^{\text{G}} \quad\forall g\\
	&P_{g,t}^{\text{G}} = p_{g,t}^{\text{G}} + \underline{P}^{\text{G}}_{g} u_{g,t}^{\text{G}} \quad\forall g\\
	&t=1,\cdots,N_{\text{T}}
	\end{align}
\end{subequations}
The dispatch changes should respect the ramp-up and ramp-down limits
\begin{subequations}
	\begin{align}
	&P_{g,t}^{\text{G}} - P_{g,t-1}^{\text{G}} \leq \overline{R}_{g}^{\text{G}}u_{g,t-1}^{\text{G}} + \underline{P}^{\text{G}}_{g} (u_{g,t}^{\text{G}} - u_{g,t-1}^{\text{G}})\forall g\\
	&P_{g,t-1}^{\text{G}} - P_{g,t}^{\text{G}} \leq \underline{R}_{g}^{\text{G}}u_{g,t-1}^{\text{G}} + \underline{P}^{\text{G}}_{g} (u_{g,t-1}^{\text{G}} - u_{g,t}^{\text{G}})\forall g\\
	& t=2,\cdots,N_{\text{T}}
	\end{align}
\end{subequations}
The commitment commands should meet the minimum-up and -down time constraints
\begin{subequations}
	\begin{align}
	& u_{g,t}^{\text{G}} - u_{g,t-1}^{\text{G}}\leq u_{g,\tau^{\text{up}}}^{\text{G}}\quad\forall g\\
	& u_{g,t-1}^{\text{G}} - u_{g,t}^{\text{G}}\leq 1-u_{g,\tau^{\text{down}}}^{\text{G}}\quad\forall g\\
	& t = 2,\cdots,N_{\text{T}}\\ 
	& \tau^{\text{up}} = t+1, \cdots, \min(t+t^{\text{up}}_{g}-1, N_{\text{T}})\\
	& \tau^{\text{down}} = t+1, \cdots, \min(t+t^{\text{down}}_{g}-1, N_{\text{T}})
	\end{align}
\end{subequations}
And for $t=1$, the minimum-up and -down time constraints read as follow
\begin{subequations}
	\begin{align}
	& u_{g,1}^{\text{G}} \leq u_{g,\tau^{\text{up}}}^{\text{G}}\quad\forall g\\
	& u_{g,1}^{\text{G}} \leq 1-u_{g,\tau^{\text{down}}}^{\text{G}}\quad\forall g\\
	& \tau^{\text{up}} = 2, \cdots, \min(t+t^{\text{up}}_{g}-1, N_{\text{T}})\\
	& \tau^{\text{down}} = 2, \cdots, \min(t+t^{\text{down}}_{g}-1, N_{\text{T}})
	\end{align}
\end{subequations}
The linearized AC power flow equations and the voltage constraints are expressed as follows \cite{trodden2014optimization}
\begin{subequations}
	\label{eq_power_flow_lin}
\begin{align}
&\begin{aligned}
&P_{i,t}^{\text{inj}}=\sum_{j\in\sigma_{b}(i)}[G_{ij}(2V_{i,t}-1)-G_{ij}(V_{i,t}+V_{j,t}-1)\\
&\qquad\qquad-B_{ij}(\theta_{i,t}-\theta_{j,t})]\quad \forall i,j,i\neq j,\forall t
\end{aligned}\\
&\begin{aligned}
&Q_{i,t}^{\text{inj}}=\sum_{j\in\sigma_{b}(i)}[B_{ij}(1-2V_{i,t}) + B_{ij}(V_{i,t}+V_{j,t}-1)\\
&\qquad\qquad-G_{ij}(\theta_{i,t}-\theta_{j,t})]\quad \forall i,j,i\neq j,\forall t\\
\end{aligned}\\
&\underline{V}\leq V_{i,t} \leq \overline{V}\quad \forall i,\forall t
\end{align}
\end{subequations}
where
\begin{subequations}
\begin{align}
&P_{i,t}^{\text{inj}}=\sum_{g\in\sigma_{g}(i)}P_{g,t}^{\text{G}}+\sum_{w\in\sigma_{w}(i)}P_{w,t}^{\text{W}}-\sum_{k\in\sigma_{d}(i)}P_{k,t}^{\text{D}}\quad \forall i,\forall t\\
&Q_{i,t}^{\text{inj}}=\sum_{g\in\sigma_{g}(i)}Q_{g,t}^{\text{G}}-\sum_{k\in\sigma_{d}(i)}Q_{k,t}^{\text{D}}\quad \forall i,\forall t
\end{align}
\end{subequations}
The linearization terms are explained in Appendix \ref{appendix_pl}.

To encode the DNN into the MILP, we will first build the feature vectors in terms of the decision variables in the UC problem. Then, we will express the trained neural networks as MILP models. As discussed before, we replace the sample index with the scheduling period index. The feature vectors for the frequency nadir and stability predictors defined in terms of the decision variables are expressed in (\ref{eq_feature_freq}) and (\ref{eq_feature_stab}), respectively. Unfortunately, $\bold{x}^{\varpi}_{t}$ contains the $\max$ operator, and $\bold{x}^{\text{f}}_{t}$ and $\bold{x}^{\text{p}}_{t}$ cannot be directly used in the encoding formulation. Thus, we introduce the supplementary variables $\nu^{\text{G}}_{g,t}$ to indicate if generator $g$ outputs the largest active power in scheduling period $t$. The re-formulations are expressed as follows
\begin{subequations}
\begin{align}
&P^{\text{G}}_{\varrho,t}-P^{\text{G}}_{g,t} \leq M(1-\nu^{\text{G}}_{g,t})\quad\forall \varrho,g\in\mathcal{N}_{\text{G}},\forall t\label{eq_max_1}\\
&\sum_{g\in\mathcal{N}_{\text{G}}}\nu^{\text{G}}_{g,t}=1\quad\forall t\label{eq_max_2}
\end{align}
\end{subequations}
where $M$ is a big number. Eq. (\ref{eq_max_1}) enforces $\nu^{\text{G}}_{g,t}$ to be zero if the output of any generators $\varrho$ is greater than $g$ during period $t$. Eq. (\ref{eq_max_2}) ensures that there should be only one largest generator during $t$. The combination of both constraints will ensure generator $g$ has the largest output if $\nu^{\text{G}}_{g,t}=1$. To further encode the magnitude information, we define another supplementary variables $\vartheta^{\varpi}_{g,t}$ and let $\vartheta^{\varpi}_{g,t}=P^{\text{G}}_{g,t}$ if $\nu^{\text{G}}_{g,t}=1$, and $\vartheta^{\varpi}_{g,t}=0$ otherwise. The corresponding constraints are expressed as follows
\begin{subequations}
	\begin{align}
	&\vartheta^{\varpi}_{g,t}-P^{\text{G}}_{g,t}\geq-M(1-\nu^{\text{G}}_{g,t})\quad\forall g\in\mathcal{N}_{\text{G}},\forall t\\
	&\vartheta^{\varpi}_{g,t}-P^{\text{G}}_{g,t}\leq M(1-\nu^{\text{G}}_{g,t})\quad\forall g\in\mathcal{N}_{\text{G}},\forall t\\
	&0\leq\vartheta^{\varpi}_{g,t}\leq M\nu^{\text{G}}_{g,t}\quad\forall g\in\mathcal{N}_{\text{G}},\forall t
	\end{align}
\end{subequations}
Finally, the feature vector for frequency nadir prediction can be re-written as follows
\begin{align}
\label{eq_dnn_X_freq_re}
\bold{x}^{\text{f}}_{t}=\left[u^{\text{G}}_{1,t},\cdots,u^{\text{G}}_{N_{\text{G}},t},\vartheta^{\varpi}_{1,t}, \cdots,\vartheta^{\varpi}_{N_{\text{G}},t}\right]
\end{align}
And the feature vector for stability prediction can be re-written as follows
\begin{align}
\label{eq_dnn_X_stab_re}
\bold{x}^{\text{p}}_{t}=\left[u^{\text{G}}_{1,t},\cdots,u^{\text{G}}_{N_{\text{G}},t},\vartheta^{\varpi}_{1,t}, \cdots,\vartheta^{\varpi}_{N_{\text{G}},t},P_{1,t}^{\text{G}},\cdots, P^{\text{G}}_{N_{\text{G}},t} \right]
\end{align}

Then, we can express the trained frequency nadir predictor $\hat{h}^{\text{f}}$ in (\ref{eq_nn_freq}) and stability predictor $\hat{h}^{\text{p}}$ in (\ref{eq_nn_stab}) as MILP models to encode the predictions of decision variables and indirectly impose frequency security and stability constraints on decision variables. Since most parts of the model are similar, we use the superscript $\alpha$ to denote the types of the neural networks, where $\alpha\in\{\text{f},\text{p}\}$. The MILP formulations read as follows
\begin{subequations}
\begin{align}
&\bold{z}^{\alpha}_{1,t}=\bold{x}^{\alpha}_{t}\bold{W}^{\alpha}_{1}+\bold{b}^{\alpha}_{1}  \quad\forall t\label{eq_dnn_milp_1}\\
&\bold{\hat{z}}^{\alpha}_{m,t}=\bold{z}^{\alpha}_{m-1,t}\bold{W}^{\alpha}_{m}+\bold{b}^{\alpha}_{m}  \quad\forall m,\forall n,\forall t\label{eq_dnn_milp_2}\\
&\bold{z}^{\alpha}_{m[n],t}\leq \bold{\hat{z}}^{\alpha}_{m[n],t}-\bold{\underline{h}}^{\alpha}_{m[n]}(1-\bold{a}^{\alpha}_{m[n],t})  \quad\forall m,\forall n,\forall t\label{eq_dnn_milp_3}\\
&\bold{z}^{\alpha}_{m[n],t}\geq \bold{\hat{z}}^{\alpha}_{m[n],t} \quad\forall m,\forall n,\forall t\label{eq_dnn_milp_4}\\
&\bold{z}^{\alpha}_{m[n],t}\leq \bold{\overline{h}}^{\alpha}_{m[n]}\bold{a}^{\alpha}_{m[n],t} \quad\forall m,\forall n,\forall t\label{eq_dnn_milp_5}\\
&\bold{z}^{\alpha}_{m[n],t}\geq 0 \quad\forall m,\forall n,\forall t\label{eq_dnn_milp_6}\\
&\bold{a}^{\alpha}_{m,t}\in\{0,1\},\alpha\in\{\text{f},\text{p}\}\label{eq_dnn_milp_7}\\
&\hat{f}_{\text{ndr},t}=\bold{z}^{\text{f}}_{N^{\text{f}}_{\text{Y}},t}\bold{W}^{\text{f}}_{N^{\text{f}}_{\text{Y}}+1}+\bold{b}^{\text{f}}_{N^{\text{f}}_{\text{Y}}+1} \quad\forall t\label{eq_dnn_milp_8}\\
&\bold{p}_{\text{stab},t}=\bold{z}^{\text{p}}_{N^{\text{p}}_{\text{Y}},t}\bold{W}^{\text{p}}_{N^{\text{p}}_{\text{Y}}+1}+\bold{b}^{\text{p}}_{N^{\text{p}}_{\text{Y}}+1}\quad\forall t\label{eq_dnn_milp_9}
\end{align}\label{eq_dnn_milp}
\end{subequations}
As shown, except for the final layers in (\ref{eq_dnn_milp_8}) and (\ref{eq_dnn_milp_9}), the rest parts of the neural networks admit the same re-formulations for both $\hat{h}^{\text{f}}$ and $\hat{h}^{\text{p}}$. A binary vector $\bold{a}^{\alpha}_{m}$ for $\alpha\in\{\text{f},\text{p}\}$ represents the activation status of ReLU at $m$th hidden layer, and $\bold{a}^{\alpha}_{m[n]}$ represents the status of the $n$th neuron at the $m$th layer. Let $[\bold{\underline{h}}^{\alpha}_{m[n]},\bold{\overline{h}}^{\alpha}_{m[n]}]$ be an interval that is large enough to contain all possible values of $\bold{\hat{z}}^{\alpha}_{m[n]}$, where $\bold{\underline{h}}^{\alpha}_{m[n]}<0$ and $\bold{\overline{h}}^{\alpha}_{m[n]}>0$. When $\bold{\hat{z}}^{\alpha}_{m[n]}$ is less than or equal to zero, constraints (\ref{eq_dnn_milp_3}) and (\ref{eq_dnn_milp_6}) will force $\bold{a}^{\alpha}_{m[n]}$ to be zero. In this case, constraints (\ref{eq_dnn_milp_5}) and (\ref{eq_dnn_milp_6}) imply that $\bold{z}^{\alpha}_{i[k]}=0$, so we have $\bold{\hat{z}}^{\alpha}_{i[k]}\leq 0 \implies \bold{a}^{\alpha}_{m[n]}=0 \implies \bold{\hat{z}}^{\alpha}_{m[n]}=0$. When $\bold{\hat{z}}^{\alpha}_{m[n]}$ is greater than zero, constraints (\ref{eq_dnn_milp_3}) and (\ref{eq_dnn_milp_6}) will force $\bold{a}^{\alpha}_{m[n]}$ to be 1. In this case, constraints (\ref{eq_dnn_milp_3}) and (\ref{eq_dnn_milp_4}) imply that $\bold{z}^{\alpha}_{m[n]}=\bold{\hat{z}}^{\alpha}_{m[n]}$, so we have $\bold{\hat{z}}^{\alpha}_{m[n]}> 0 \implies \bold{a}^{\alpha}_{m[n]}=1 \implies \bold{z}^{\alpha}_{m[n]}=\bold{\hat{z}}^{\alpha}_{m[n]}$. Obviously, this formulation contains no approximation of the original model. In addition, this is the tightest possible formulation with respect to its LP relaxation if no future information about $\bold{\hat{z}}^{\alpha}_{m[n]}$ is revealed \cite{Anderson2019}.

Finally, we impose the frequency security constraint and stability constraint on the predictions
\begin{subequations}
\label{eq_enforce_fc}
\begin{align}
&\hat{f}_{\text{ndr},t} \geq f_{\text{UFLS}}\quad\forall t\label{eq_enforce_fc_1}\\
&(\bold{p}_{\text{stab},t})_{[2]} \geq (\bold{p}_{\text{stab},t})_{[1]} + \varepsilon_{\text{stab},t}\quad\forall t\label{eq_enforce_fc_2}
\end{align}
\end{subequations}
where $\varepsilon_{\text{stab},t}$ is the supplementary variable to relax the predicted stability condition. Since only the steady-state information is used for dynamic stability prediction, the accuracy is relatively low. Therefore, strictly enforcing the predicted stability condition requires significant searches for feasible solutions and is not computationally efficient. Essentially, the stability predictor tries to filter out as many samples as possible that the frequency nadir predictor does not encounter. The relaxed formulation can sufficiently minimize this risk.

The two-segment cost function, consisting of the fixed and marginal costs, is employed. The scheduling objective is to minimize the total operational cost, expressed as follows
\begin{subequations}
	\label{eq_obj}
	\begin{align}
	\quad {\text{min}} \quad & \sum_{t=1}^{N_\text{T}}\sum_{g=1}^{N_{\text{G}}}\left[ \lambda^{\text{M}}_{g}p_{g,t}^{\text{G}}+\lambda^{\text{F}}_{g}u^{\text{G}}_{g,t}\right] \label{eq_obj_op}\\
	&+ \sum_{t=2}^{N_\text{T}-1}\sum_{g=1}^{N_{\text{G}}}\lambda^{\text{S}}_{g}w^{\text{G}}_{g,t}\label{eq_obj_sp}\\
	&+\sum_{t=1}^{N_\text{T}} \varGamma\varepsilon_{\text{stab},t}\label{eq_obj_stab_relax}
	\end{align}
\end{subequations}
where $p_{g,t}^{\text{G}}$ denotes the incremental outputs of SG $g$, $\lambda^{\text{F}}$ and $\lambda^{\text{M}}$ denote the fixed and marginal costs, respectively. Terms (\ref{eq_obj_op}) and (\ref{eq_obj_sp}) represent the fuel and start-up costs of SGs, respectively. Terms (\ref{eq_obj_stab_relax}) represents the stability risk with a weighting factor $\varGamma$. It is worth noting that reformulation of the start-up cost has already been carried out in (\ref{eq_obj_sp}), where $w^{\text{G}}_{g,t}$ is the slack binary variable. In addition, $w^{\text{G}}_{g,t}$ and $u^{\text{G}}_{g,t}$ are subject to the following constraints
\begin{equation}
\begin{aligned}
&w^{\text{G}}_{g,t}\geq 0,w^{\text{G}}_{g,t}\geq u^{\text{G}}_{g,t}-u^{\text{G}}_{g,t-1}\quad\forall g, t=2,\cdots,N_{\text{T}}
\end{aligned}
\end{equation}

\section{Case Study}\label{sec_case}
\begin{figure}[h]
	\centering
	\includegraphics[scale=0.25]{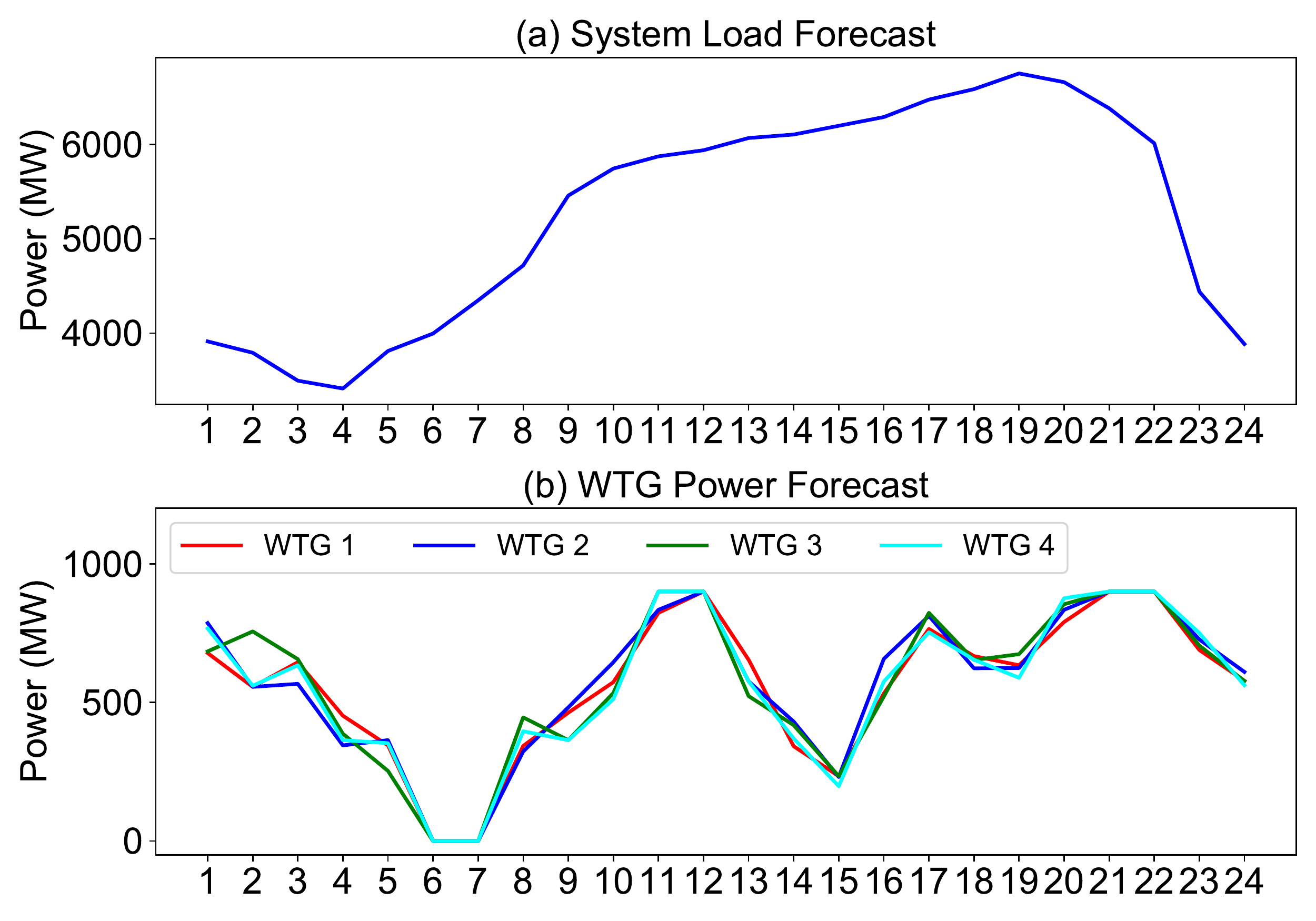}
	\vspace{-0.15in}
	\caption{(a) System load forecast. (b) WTG power forecast.}
	\label{fig_Forecast}
\end{figure}

We use the IEEE 39-bus system to demonstrate the framework. The PSS/E software has been widely used in the power industry for dynamic security assessment in daily operations and thus is chosen for labeling the samples. We use full-scale models for the dynamic simulation during the labeling process: \texttt{GENTPJU1} for the synchronous machine; \texttt{IEEEX1} for the excitation system; \texttt{IEESGO} for the turbine-governor; \texttt{PSS2A} for the power system stabilizer. We assume there are four WTGs installed at Buses 2, 10, 20, 25, respectively. Each WTG is an equivalent representation of 600 1.5MW realistic Type-3 WTGs. Standard WTG and corresponding control modules are employed: \texttt{WT3G2} for the doubly-fed induction machine, \texttt{WT3E1} for the electric control, \texttt{WT3T1} for the mechanical model, \texttt{WT3P1} for the pitch control. Details of these models can be found in \cite{siemens2009pss}. The forecast aggregated load and wind power are plotted in Fig. \ref{fig_Forecast}. The aggregated load is distributed to each bus based on the ratio of demand in the power flow data.

%\footnote{The parameters of these models can be found in \url{https://github.com/whoiszyc/AL_set_learning}}
\begin{figure}[h]
	\centering
	\includegraphics[scale=0.27]{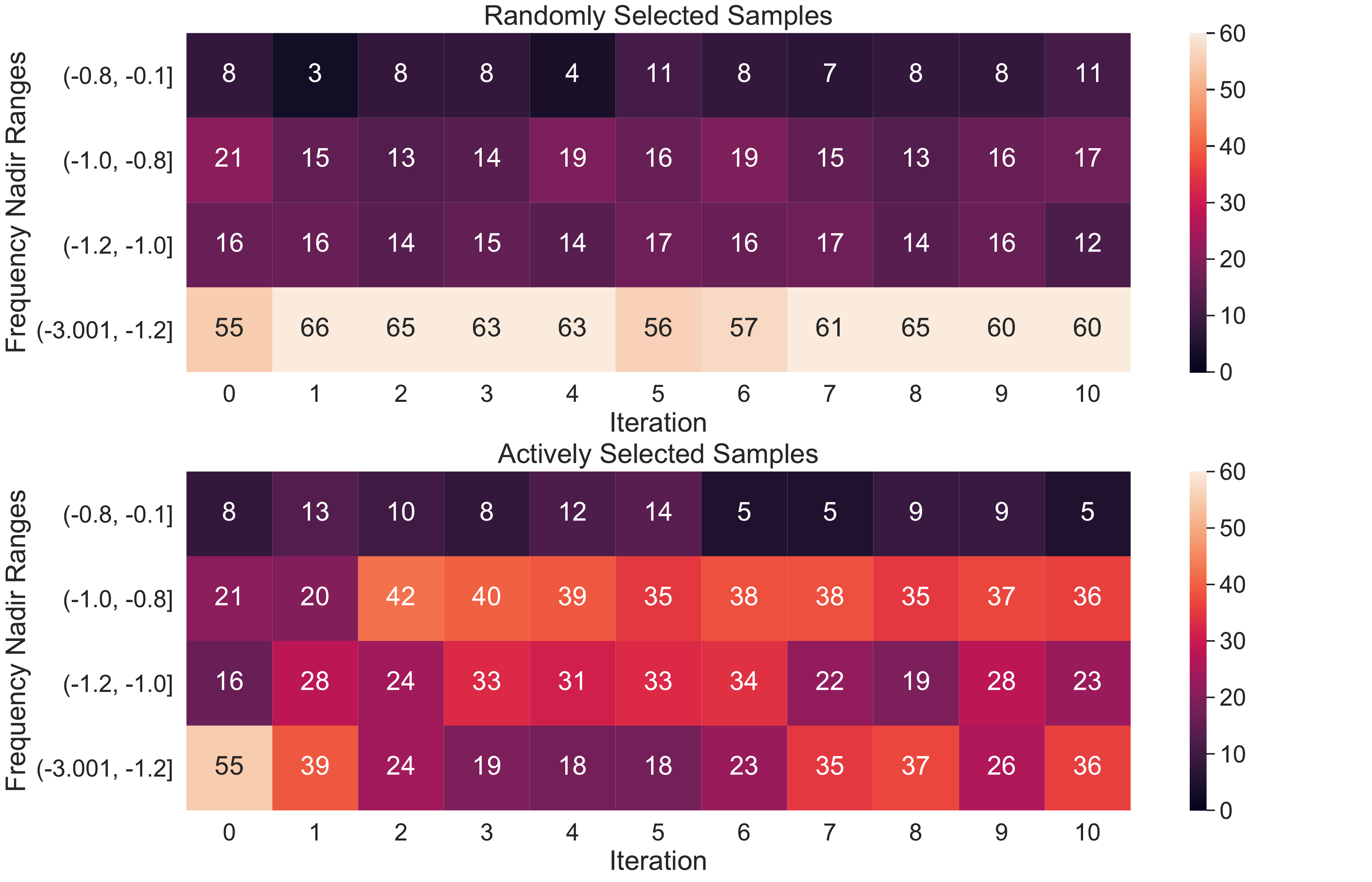}
	\vspace{-0.15in}
	\caption{Samples selected by active and random strategies. The number in each box represents how many samples whose label values are within the interval are selected in this iteration. The active method selects more samples within the region of interest, that is, close to the UFLS threshold.}
	\label{fig_al_data}
\end{figure}

\subsection{Post-Processing for Sample Labeling}\label{sec_sub_process}
To learn the underlying frequency characteristics of the system, it is important that turbine-governor modes are dominant in the collected dynamics responses. Generator tripping is the ideal event and has been used in most FCUC studies. 
However, the post-disturbance trajectories can still exhibit different unstable scenarios and possible numerical issues. Post-processing is essential after each simulation to create correct and consistent labels. It is worth mentioning that the \emph{minimum bus frequency deviation nadir} is used to label the samples. The bus frequencies are considered to be the most robust measure to avoid UFLS events since the UFLS relay is installed at each load bus. Let $T_0$ be the time instant when the disturbance is imposed, $T_1$ be the time instant when the turbine-governor response reaches the steady-state, and $T_2$ be the end instant of the simulation. The following steps are performed for post-processing
	\begin{enumerate}
		\item If the network does not converge during the simulation, mark the sample as unstable. This information indicates that the system does not have a power flow solution after the disturbance.
		\item If the simulation successfully completes, check the maximum rotor angle difference of the system. If the maximum rotor angle difference is greater than 360 degrees, mark the sample as unstable. This indicates the system is transient unstable.
		\item If the simulation is transient stable, check the minimum voltage nadir of all buses. If the minimum voltage nadir is below 0.65 per unit, mark the sample as unstable. The reason is that when the system voltage is too low, many existing models cannot represent the true behaviors of the system and the simulated trajectories are no longer accurate.
		\item Check the distances between the maximum and minimum rotor angles for all generators in time windows $T_1+\Delta T$ and $T_2-\Delta T$, denoted as $D_1$ and $D_2$, respectively. If $D_2 > D_1$, the system is small-signal unstable.
		\item If the sample is stable, obtain simulated trajectory data of all bus frequencies. Remove the first 0.1 second data after the disturbance for all bus frequencies and get the frequency nadir as the label. The reason to remove the data in that time frame is to filter out the numerical issue of washout filter-based bus frequency computation. A detailed discussion of this issue can be found in \cite{Milano2017}.
\end{enumerate}

\subsection{Region-of-Interest Active Sampling and DNN Training}
We first generate 5000 power injection samples as described in (\ref{eq_full_injection}) with negligible computation efforts. For generators, the active power injections are uniformly sampled over the lower and upper bounds. We assume SGs have adequate reactive power capacity, and WTGs are controlled with a unity power factor. Therefore, we do not need to sample the reactive power of all generators. For loads, both active and reactive power are sampled based on the Gaussian distribution. The means are equal to the demand values in the power flow data, and deviations are equal to 30\% of their nominal values. It is worth noting that the iteration number and the batch sampling number in each iteration are two important hyperparameters in the active sampling framework. A trial-and-error process and empirical studies will be needed to determine the best parameters. As we know, training performance is not proportional to the sample size and will reach a limit after a specific size of samples. Therefore, in this paper, we will set the target training performance, which is 0.9 $R^2$ score, and try to use a minimal number of samples to achieve the target. 

With all these unlabeled samples, we employ the \emph{actively-select-then-label} strategy described in Section \ref{sec_al_sampling} to generate the training dataset. For comparison, another dataset is generated using the \emph{randomly-select-then-label} procedure. In each iteration, 100 samples will be selected for PSS/E to label. We divide the nadir values of frequency deviation into four intervals: $[-3,-1.2]$, $[-1.2,-1.0]$, $[-1.0,-0.8]$, $[-0.8,-0.1]$. The regions of interest are  $[-1.2,-1.0]$ and $[-1.0,-0.8]$ since they are closer to the UFLS threshold. The heatmap of the selected samples with both active and random methods is shown in Fig. \ref{fig_al_data}. The number in each box represents how many samples whose label values are within the interval are selected in this iteration. As shown, the active method selects more samples within the region of interest. In addition, if we use random sampling, more than 50\% of the samples are unstable and not available for frequency nadir predictor training. This number will reduce to around 25\% if the active sampling strategy is employed and improve labeling efficiency.

We use \emph{actively-select-then-label} and \emph{randomly-select-then-label} strategies to train two frequency nadir predictor, respectively. We demonstrate the out-of-sample validation under both region-of-interest and full-range sample sets.  To do this, we first generate another 5000 randomly selected samples, label these samples using simulations, and select stable ones based on the post-processing criterion in Section \ref{sec_sub_process}. To validate the region-of-interest prediction, we select post-processed samples whose label values are within the most concerning ranges, which is [-1.2, 0] Hz. For full-range validation, we can directly plug in all post-processed samples. The following metrics are used to demonstrate the prediction accuracy: (1) maximum error (MAX-E), (2) median absolute error (MED-E), (3) mean absolute error (MEA-E), and (4) $R^{2}$ score, which is defined as follows
\begin{align}
R^{2}(f_{\text{ndr},s},\hat{f}_{\text{ndr},s})=1-\frac{\sum_{s=1}^{N_{\text{S}}}(f_{\text{ndr},s}-\hat{f}_{\text{ndr},s})^{2}}
{\sum_{s=1}^{N_{\text{S}}}(f_{\text{ndr},s}-\overline{f}_{\text{ndr}})^{2}}
\end{align}
where $\overline{f}_{\text{ndr}}$ is the mean of all actual labels. $R^{2}$ score gives some information about the goodness of model fitting. In regression, the $R^{2}$ coefficient of determination is a statistical measure of how well the regression predictions approximate the real data points. An $R^{2}$ of 1 indicates that the regression predictions perfectly fit the data. 

\begin{figure}[h]
	\centering
	\includegraphics[scale=0.35]{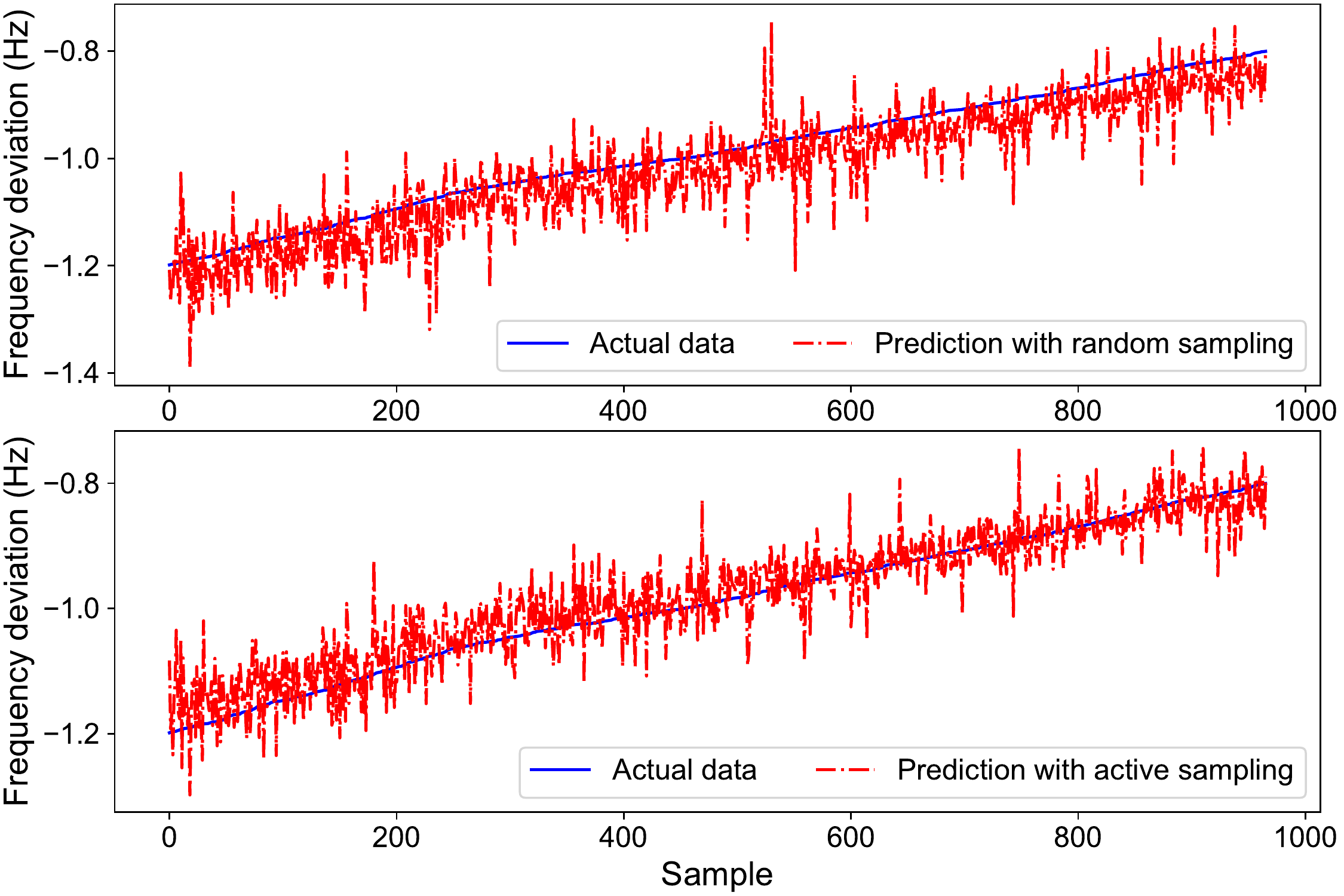}
	\caption{Region-of-interest out-of-sample prediction. The upper plot is the predictor trained by randomly sampled data, and lower plot is the predictor trained by actively sampled data.}
	\label{fig_predict_freq}
\end{figure}

\begin{figure}[h]
	\centering
	\includegraphics[scale=0.35]{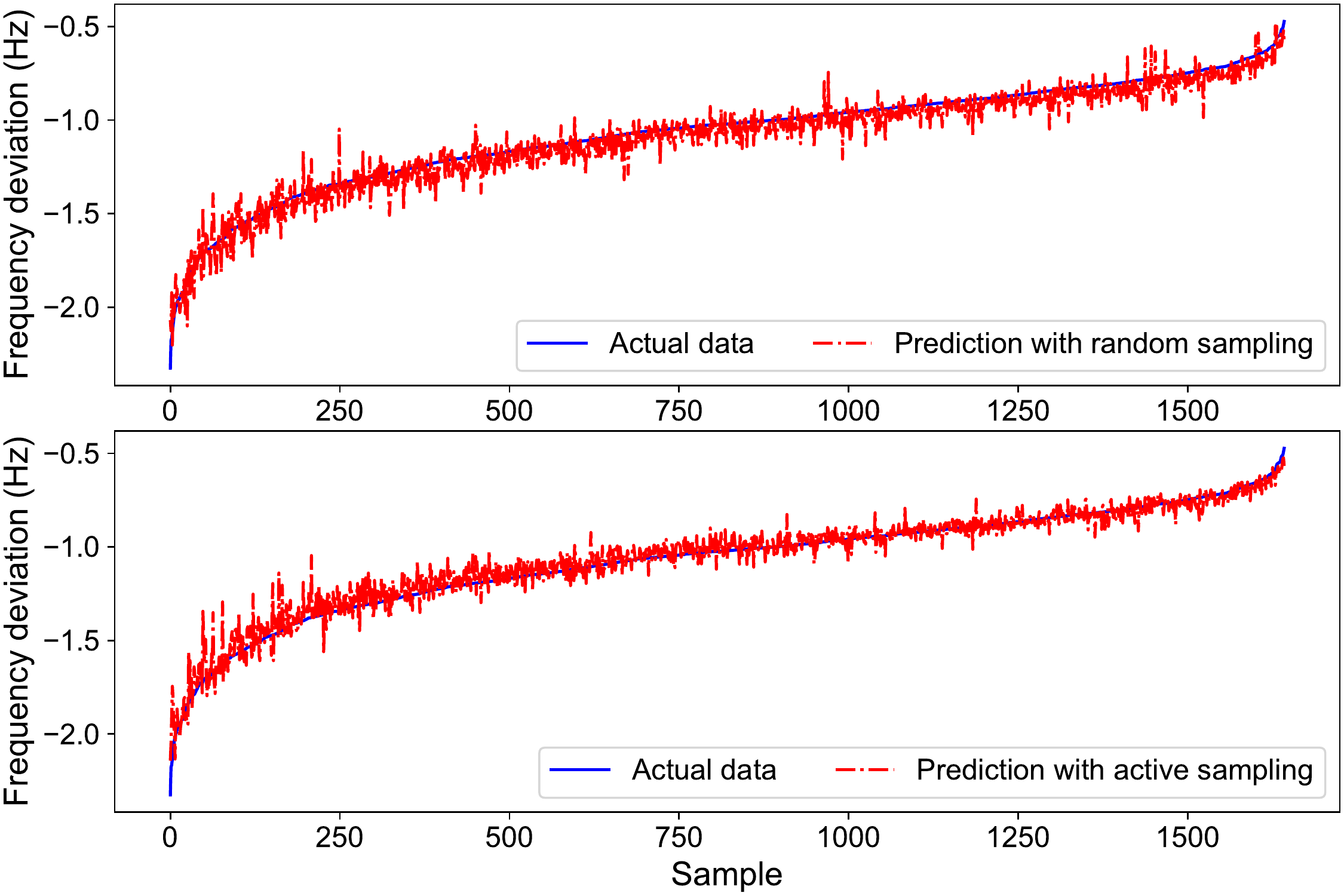}
	\caption{Full-range out-of-sample prediction. The upper plot is the predictor trained by randomly sampled data, and lower plot is the predictor trained by actively sampled data.}
	\label{fig_predict_freq_full}
\end{figure}

\begin{figure}[h]
	\centering
	\includegraphics[scale=0.25]{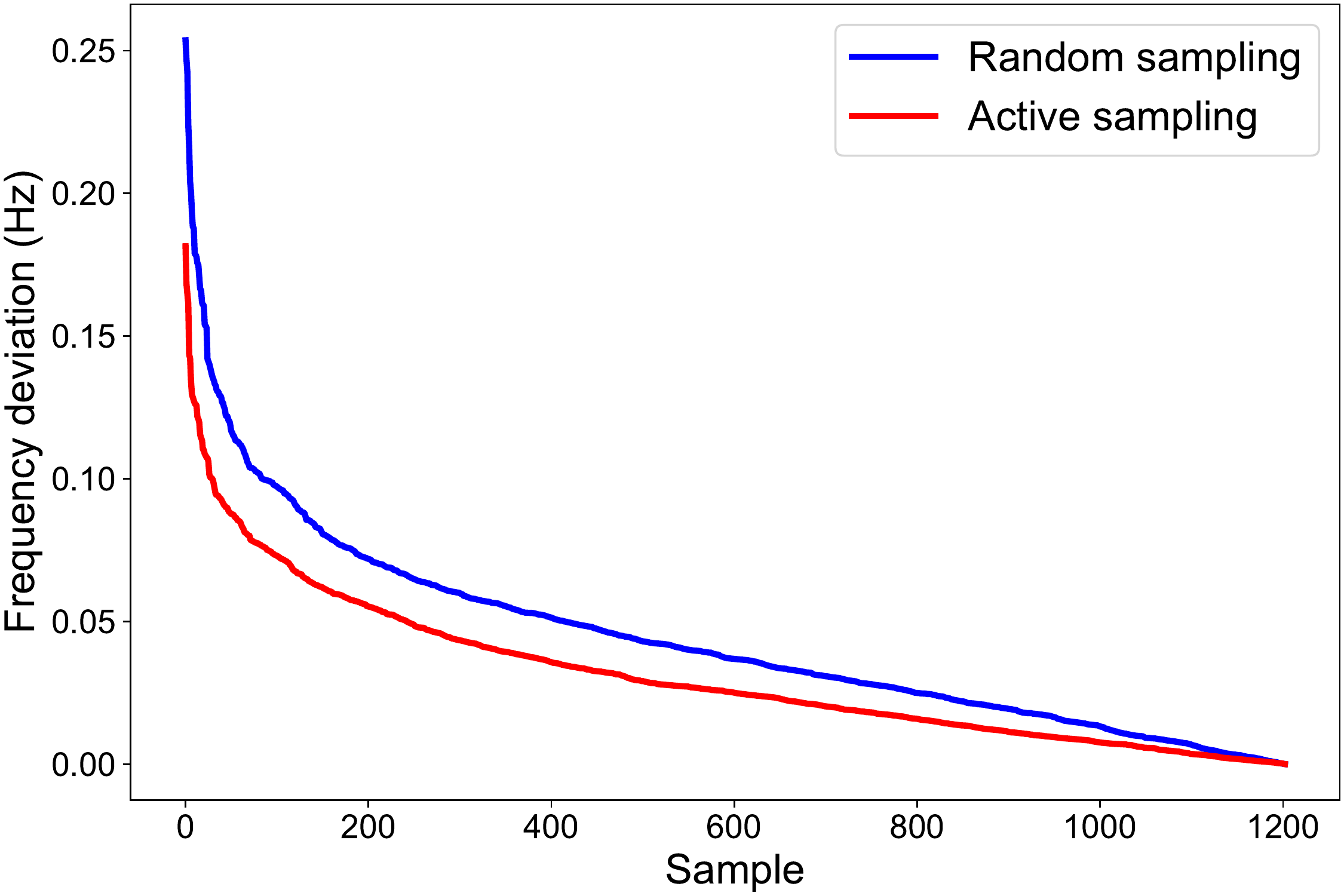}
	\caption{Region-of-interest out-of-sample absolute prediction errors. All absolute prediction errors are re-organized in descending order.}
	\label{fig_fig_predict_error}
\end{figure}

\begin{figure}[h]
	\centering
	\includegraphics[scale=0.25]{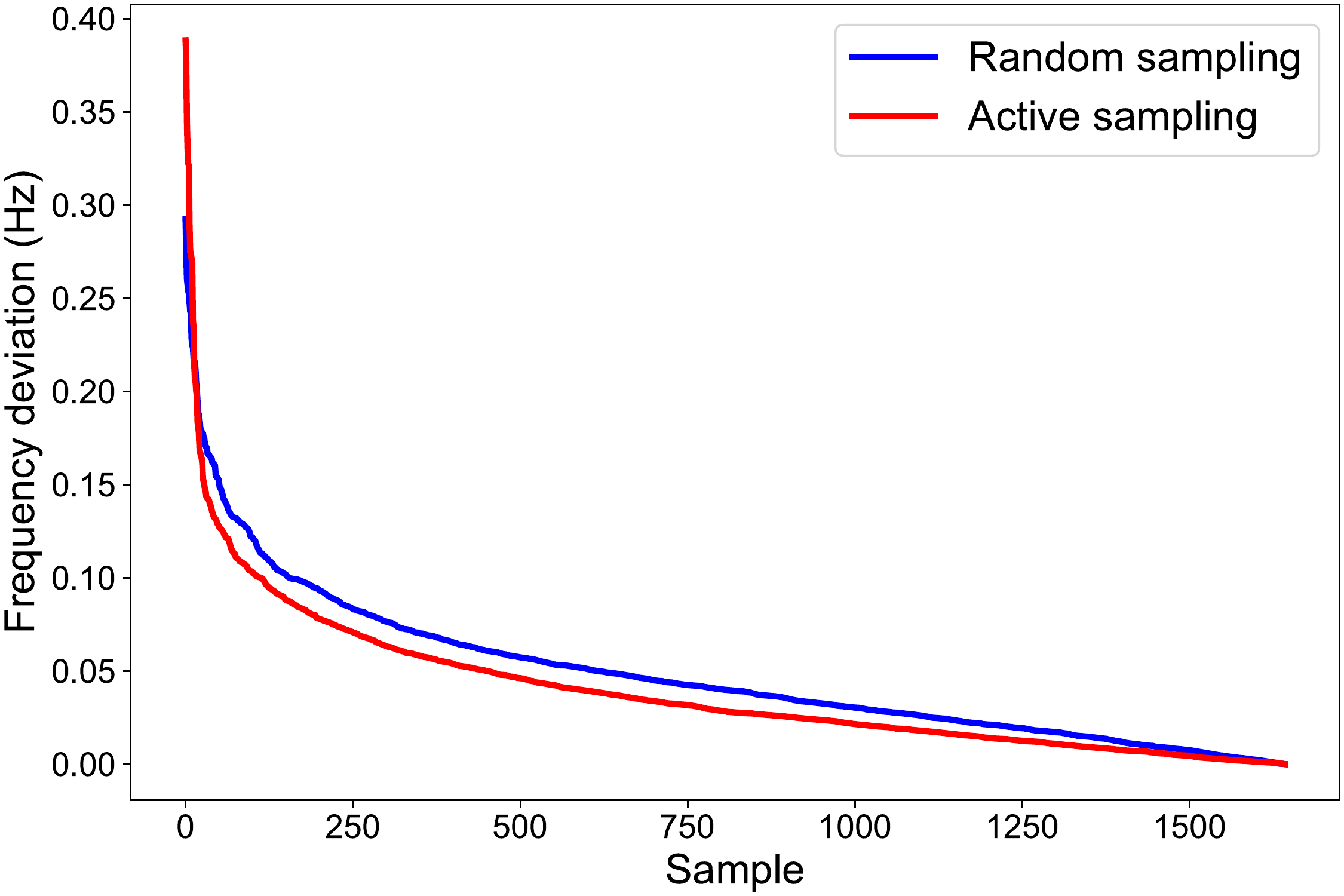}
	\caption{Full-range out-of-sample absolute prediction errors. All absolute prediction errors are re-organized in descending order.}
	\label{fig_predict_error_full}
\end{figure}

The region-of-interest and full-range \emph{out-of-sample} prediction results are shown in Fig. \ref{fig_predict_freq} and Fig. \ref{fig_predict_freq_full}, respectively. Their statistical information is concluded in Table \ref{tab_al_validation} and \ref{tab_al_validation_full}, respectively. In both cases, the predictor trained by actively sampled data is compared with the one trained by randomly sampled data. As we expected, in region-of-interest prediction, the predictor trained by actively sampled data significantly outperforms the one trained by randomly sampled data as shown in Table \ref{tab_al_validation}. This superiority can also be seen in Fig. \ref{fig_fig_predict_error}, where all prediction errors are re-organized in descending order. As for the full-range prediction, the predictor trained by actively sampled data slightly outperforms the predictor built by randomly sampled data in terms of MAX-E, MEA-E, and $R^{2}$, as shown in Table \ref{tab_al_validation_full} and Fig. \ref{fig_predict_error_full}. But the predictor trained by actively sampled data admits a larger MAX-E. It is reasonable since actively sampled data may lack samples outside the region-of-interest for generalizing the predictor. Fortunately, this large error occurs when the true frequency deviation is far from the UFLS threshold and will have little impact on the UC since it will induce a binding constraint anyway. We have also compared the out-of-sample prediction results between the DNN, linear regression, support vector regression (SVR), and random forest (RF) in Table \ref{tab_ml_comp}. As we can see, the DNN model outperforms other models in most of the metrics, especially in MAX-E.

In this paper, we use the same structure for the frequency and stability predictors, that is, $m=1$ and $n=[256]$ (number of neutrons in each hidden layer). For the neural network used for active sampling, $m=2$ and and $n=[512, 128]$.

\begin{table}[H]
	\caption{Region-of-Interest Out-of-Sample Prediction}
	\centering
	\begin{tabular}{lclclclclclclcl}
		\toprule 
		Case & MAX-E & MED-E  & MEA-E & $R^{2}$\\
		\midrule
		Active & 0.1814 (Hz) & 0.0250 (Hz) & 0.0314 (Hz) & 0.9237 \\
		Random & 0.2536 (Hz) & 0.0368 (Hz) & 0.0445 (Hz)  & 0.8536 \\
		\bottomrule
	\end{tabular}
	\label{tab_al_validation}
\end{table}
\begin{table}[H]
	\caption{Full-range Out-of-Sample Prediction}
	\centering
	\begin{tabular}{lclclclclclclcl}
		\toprule 
		Case & MAX-E & MED-E  & MEA-E & $R^{2}$\\
		\midrule
		Active & 0.3883 (Hz) & 0.0277 (Hz) & 0.0395 (Hz) & 0.9583 \\
		Random & 0.2925 (Hz) & 0.0393 (Hz) & 0.0486 (Hz)  & 0.9477 \\
		\bottomrule
	\end{tabular}
	\label{tab_al_validation_full}
\end{table}
\begin{table}[H]
	\caption{Full-range Out-of-Sample Prediction with Different Methods}
	\centering
	\begin{tabular}{lclclclclclclcl}
		\toprule 
		Case & MAX-E & MED-E  & MEA-E & $R^{2}$\\
		\midrule
		DNN & 0.3883 (Hz) & 0.0277 (Hz) & 0.0395 (Hz) & 0.9583 \\
		Linear & 0.4282 (Hz) & 0.0334 (Hz) & 0.0411 (Hz)  & 0.9605 \\
		SVR & 0.5691 (Hz) & 0.0704 (Hz) & 0.0859 (Hz)  & 0.8413 \\
		RF & 0.8070 (Hz) & 0.0411 (Hz) & 0.0620 (Hz)  & 0.8882 \\
		\bottomrule
	\end{tabular}
	\label{tab_ml_comp}
\end{table}

\subsection{FCUC Results}
\begin{figure}[h]
	\centering
	\includegraphics[scale=0.3]{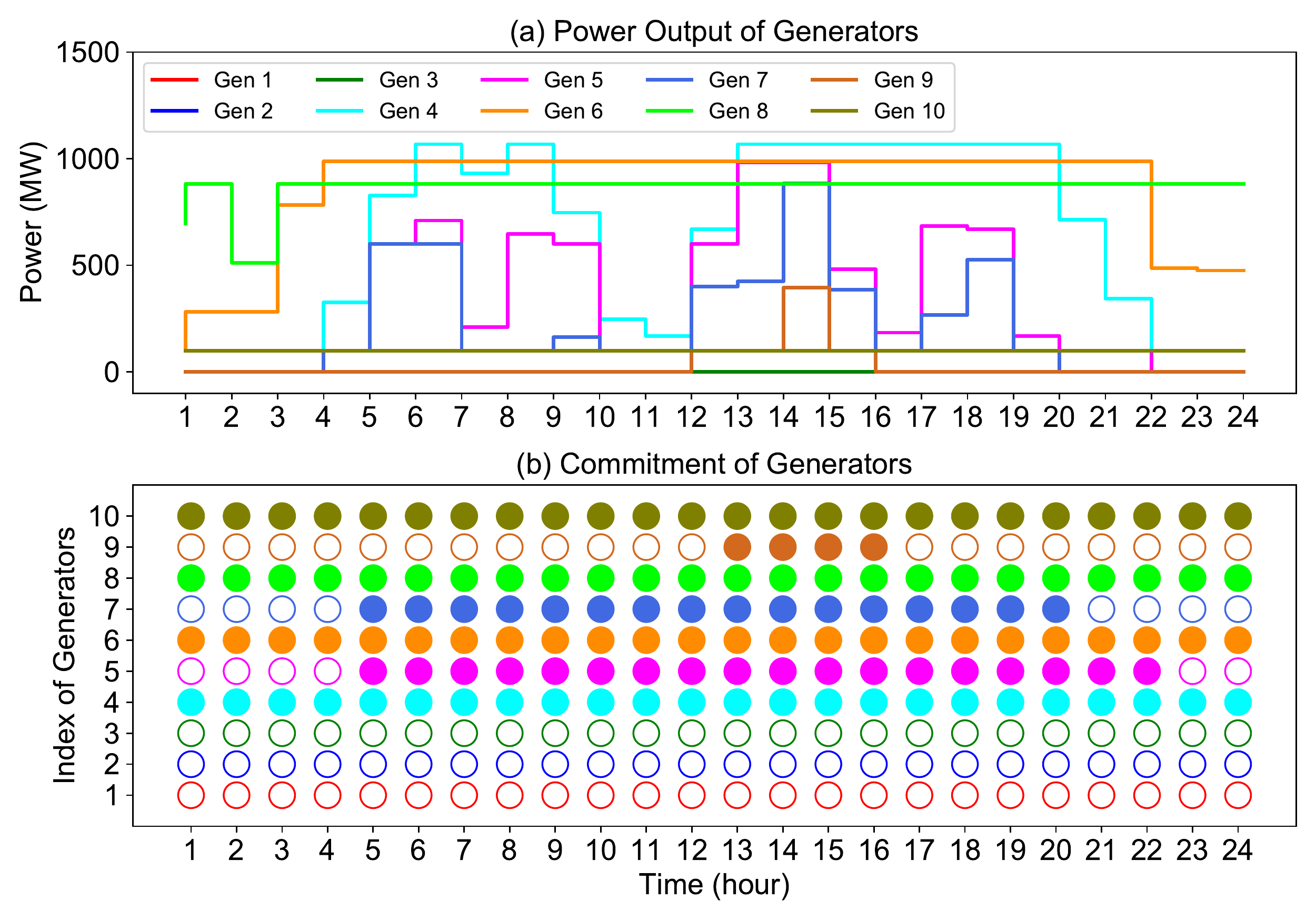}
	\vspace{-0.15in}
	\caption{UC results using ordinary formulations. (a) Power outputs of SGs. (b) Commitment of SGs.}
	\label{fig_UC_normal}
\end{figure}

\begin{figure}[h]
	\centering
	\includegraphics[scale=0.3]{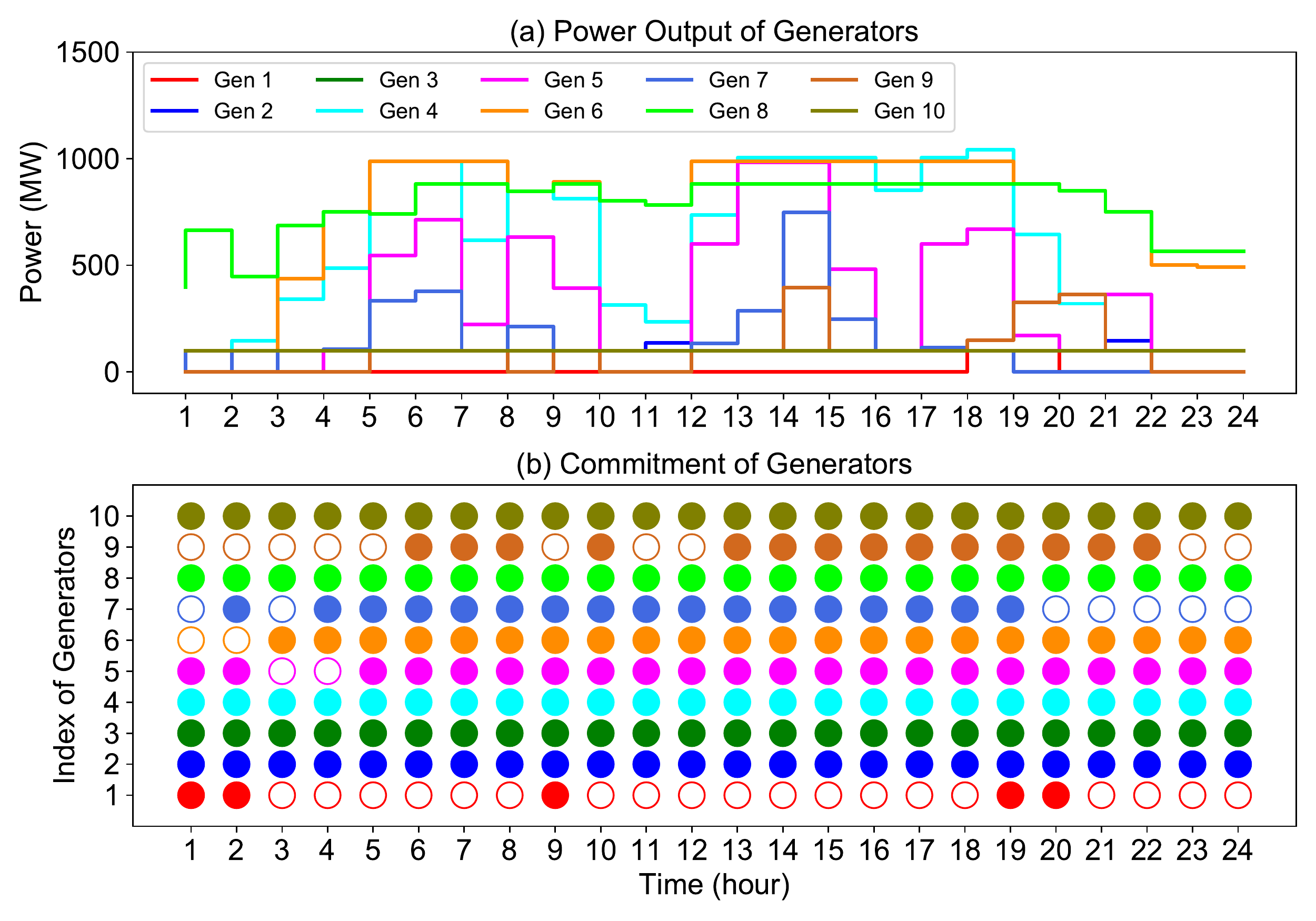}
	\vspace{-0.15in}
	\caption{UC results with frequency trajectory constraints. (a) Power outputs of SGs. (b) Commitment of SGs.}
	\label{fig_UC_FC}
\end{figure}

The voltages are allowed to variate between 0.8 and 1.2 p.u. The unit commitment results of ordinary formulations (Eqs. (\ref{eq_gen_dispatch})-(\ref{eq_power_flow_lin})) and (\ref{eq_obj}) are shown in Fig. \ref{fig_UC_normal}. While, the results of FCUC formulations (Eqs. (\ref{eq_gen_dispatch})-(\ref{eq_obj})) are shown in Fig. \ref{fig_UC_FC}, where the frequency security limit is set to be 59 Hz. A stability predictor with 78\% accuracy is trained and incorporated. As we can see, FCUC formulation enforces more SGs to be committed such that adequate responsive active and reactive power is online to ensure both stability and dynamic frequency security. The FCUC also makes the outputs of different SGs to be more equalized to reduce the risk of stability and security under the worst-case contingency. Particularly, the system under ordinary dispatch strategy is vulnerable to the $N-1$ generator outage event during nights. These are periods when the wind power is abundant and demand is deficient, resulting in less committed SGs. For example, in the scheduling periods 23 and 24, ordinary UC only schedules four SGs shown in Fig. \ref{fig_UC_normal}. In contrast, FCUC formulation commits seven SGs illustrated in Fig. \ref{fig_UC_FC}. 

\begin{figure}[h]
	\centering
	\includegraphics[scale=0.25]{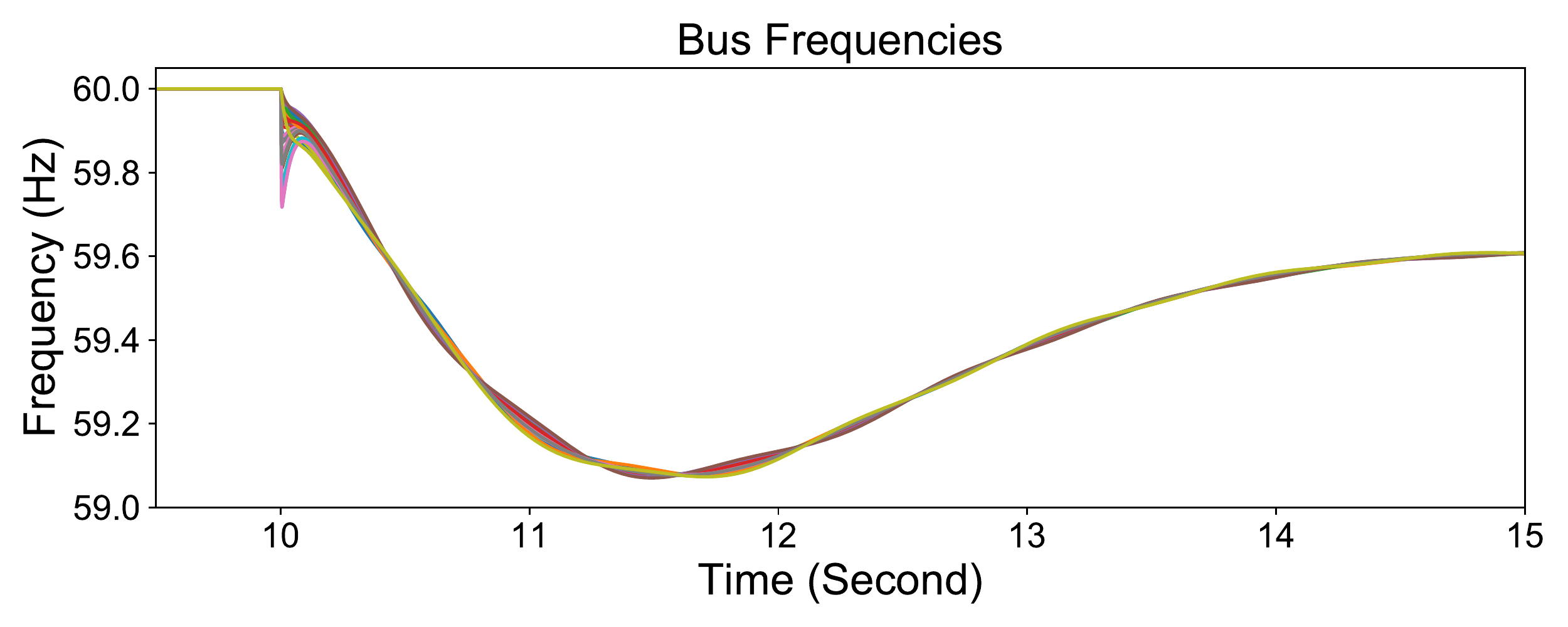}
	\vspace{-0.15in}
	\caption{Frequency response under the FCUC dispatch and worst-case contingency during the scheduling period 24. }
	\label{fig_dyn_freq}
\end{figure}

\begin{figure}[h]
	\centering
	\includegraphics[scale=0.25]{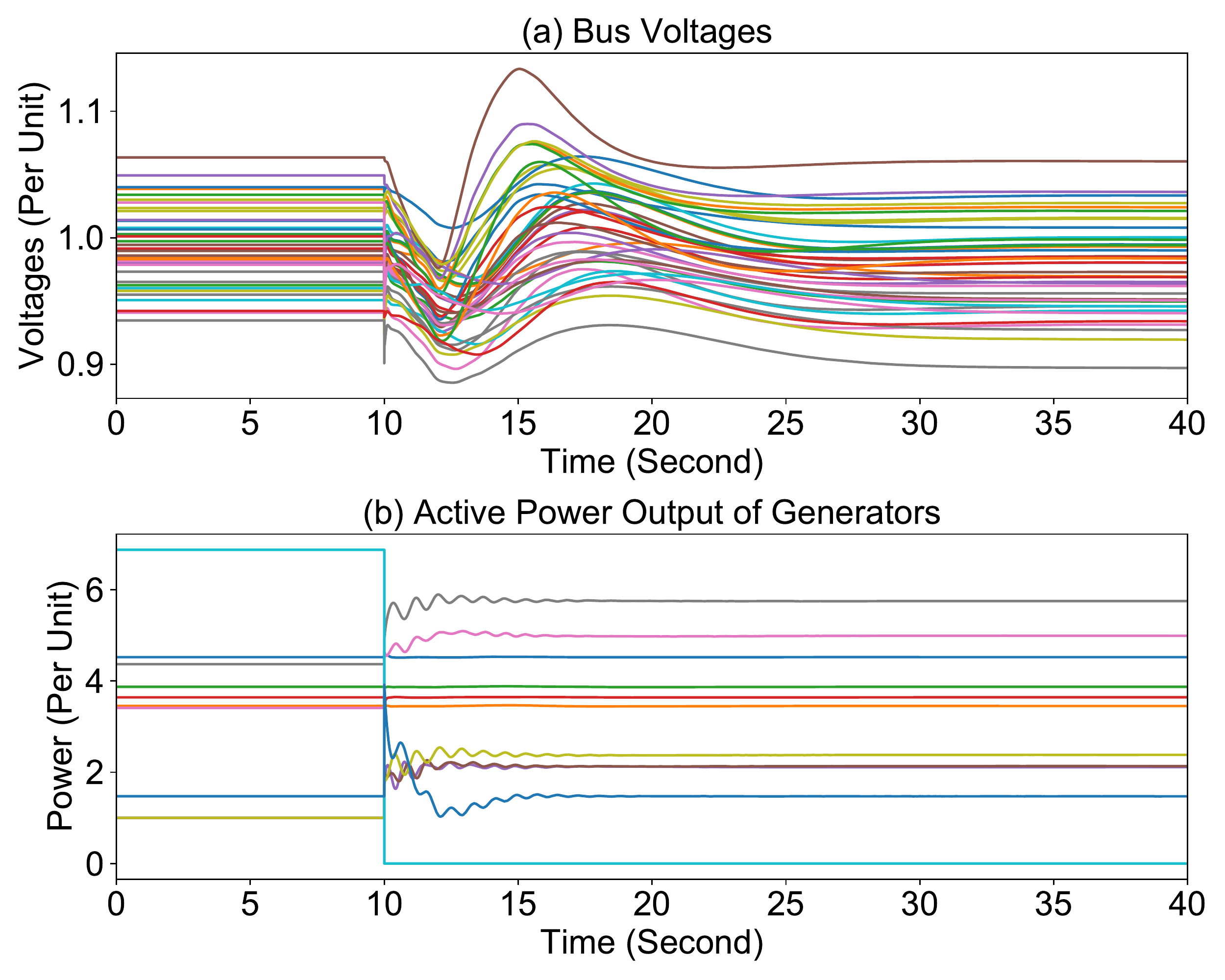}
	\vspace{-0.15in}
	\caption{System dynamic response under the FCUC dispatch and worst-case contingency during the scheduling period 24. (a) Bus voltages. (b) Generator active power outputs.}
	\label{fig_dyn_other}
\end{figure}

We assume the worst-case contingency takes place in period 24, and the generator outputting the largest power is tripped. The system under the ordinary UC schedules does not have a power flow solution after the disturbance. While, the system frequency responses following the FCUC dispatch are shown in Fig. \ref{fig_dyn_freq}, which satisfies the frequency trajectory constraint. The voltage dynamics and active power outputs of generators are plotted in Fig. \ref{fig_dyn_other} (a) and (b), respectively. Fig. \ref{fig_dyn_other} (a) indicates that the FCUC dispatch also satisfy the voltage limits. Fig. \ref{fig_dyn_other} (b) illustrates the generator tripping and responses from other generators except for the WTGs. 

The systematic verification using PSS/E is shown in Table \ref{tab_nadir}. The first column is the predicted deviated frequency nadir by the FCUC. The second column is the corresponding PSS/E verification. The third column is the PSS/E simulation of the ordinary UC dispatch. The term "NA" denotes instability cases. As we can see, ordinary UC dispatch is severely insecure as most dispatched conditions cannot withstand the worst-case contingency. Meanwhile, the FCUC significantly improves stability and security with satisfactory prediction accuracy.
\begin{table}[H]
	\caption{Comparison of deviated frequency nadirs in Hz from FCUC prediction and PSS/E verification}
	\centering
	\begin{tabular}{lclclclclclclclcl}
		\toprule 
		$\bold{Period}$ & Pred. & FCUC & UC  & $\bold{Period}$ & Pred. & FCUC & UC\\
		\midrule
		$\mathbf{1}$ & -0.75 & -0.61 & NA & $\mathbf{13}$ & -0.99 & -0.99 & NA  \\
		$\mathbf{2}$ & -0.90 & -0.84 &  NA & $\mathbf{14}$ & -1.00 & -0.98 & NA  \\
		$\mathbf{3}$ & -1.00 & -0.72 &  NA & $\mathbf{15}$ & -1.00 & -1.04 & NA    \\ 
		$\mathbf{4}$ & -1.00 & -0.93 &  NA & $\mathbf{16}$ & -1.00 & -0.99 & NA    \\
		$\mathbf{5}$ & -0.94 & -0.91 &  -1.66 & $\mathbf{17}$ & -0.99 & -0.99 & NA     \\
		$\mathbf{6}$ & -0.98 & NA &  NA & $\mathbf{18}$ & -1.00 & -0.98 & NA    \\
		$\mathbf{7}$ & -0.98 & NA &  NA & $\mathbf{19}$ & -1.00 & -0.98 & NA    \\
		$\mathbf{8}$ & -0.99 & -0.96 &  NA & $\mathbf{20}$ & -0.98 & -0.99 & NA    \\
		$\mathbf{9}$ & -1.00 & -0.99 &  NA & $\mathbf{21}$ & -1.00 & -1.00 & NA   \\
		$\mathbf{10}$ & -0.89 & -0.88 &  NA & $\mathbf{22}$ & -0.89 & -0.87 & NA   \\
		$\mathbf{11}$ & -1.00 & -1.02 &  NA & $\mathbf{23}$ & -1.00 & -0.75 & NA   \\
		$\mathbf{12}$ & -0.98 & -1.01 &  NA & $\mathbf{24}$ & -1.00 & -0.76 & NA   \\
			\bottomrule
	\end{tabular}
	\label{tab_nadir}
\end{table}

\section{Discussions}\label{sec_discuss}
\subsection{On Different System Frequency Response Models in FCUC}\label{sec_discuss_sub_sfr}
\begin{figure}[h]
	\centering
	\includegraphics[scale=0.3]{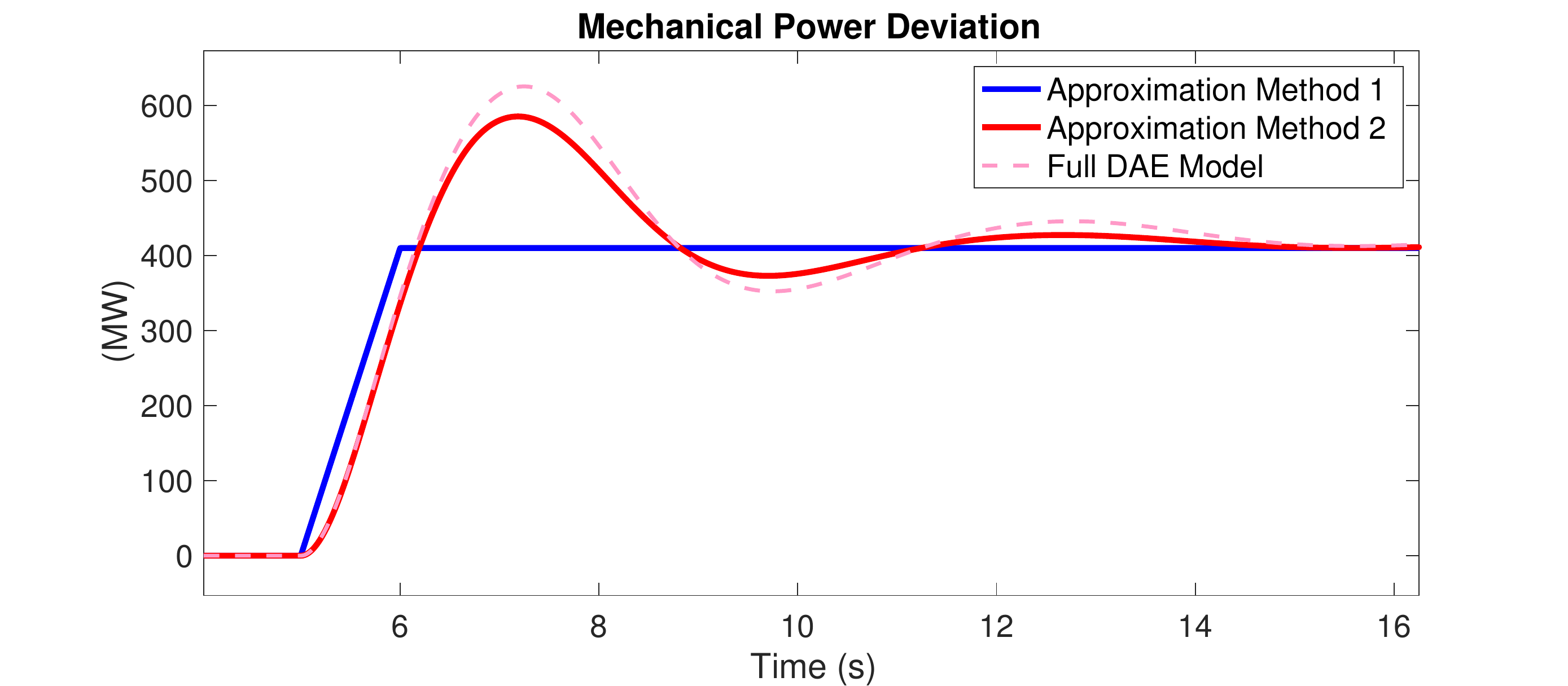}
	\caption{Mechanical power captured by two typical model-based approximation methods. Our approach can directly encode the mechanical power dynamics of the full DAE model.}
	\label{fig_dyn_comp_pm}
\end{figure}

\begin{figure}[h]
	\centering
	\includegraphics[scale=0.3]{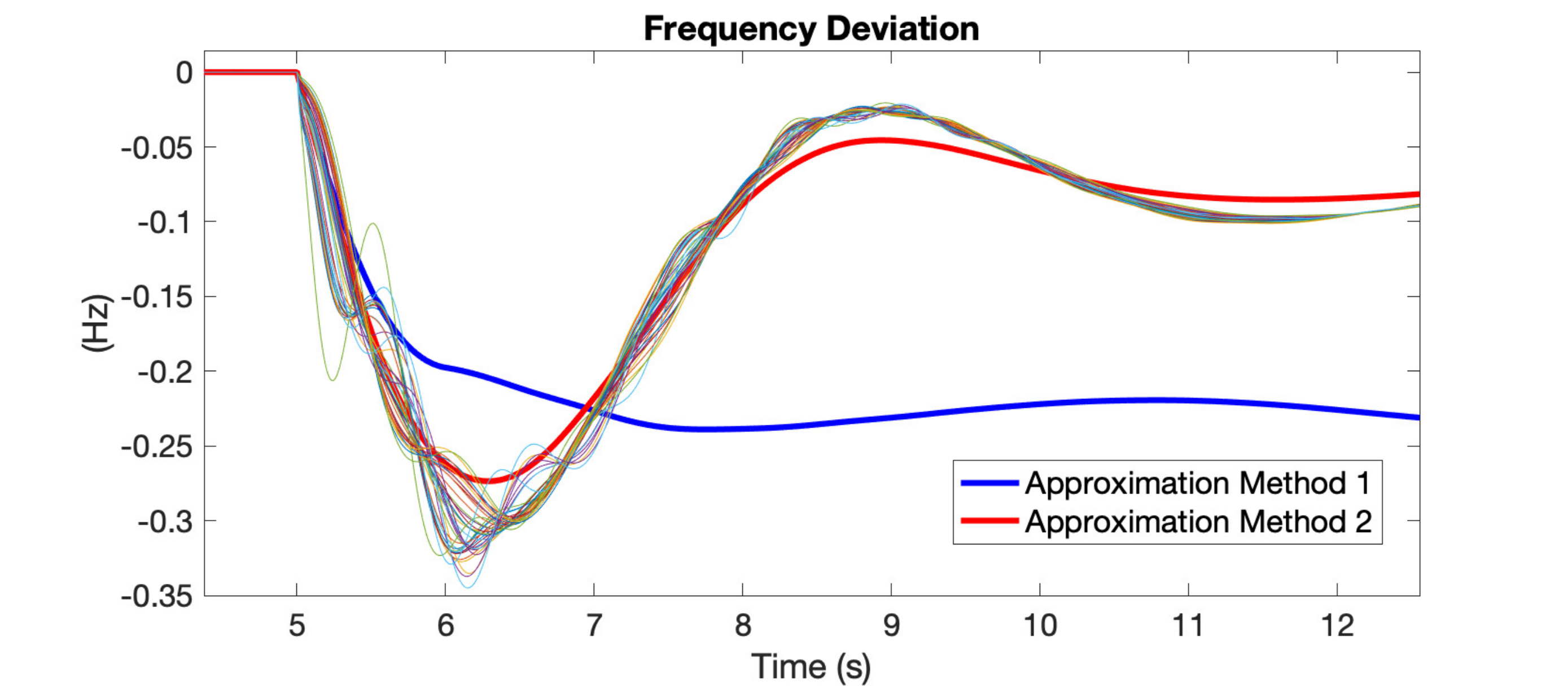}
	\caption{Frequency responses captured by two typical existing model-based approximation methods. Our approach can directly encode the bus frequency responses of the full DAE model.}
	\label{fig_dyn_comp_f}
\end{figure}

Essentially, it is the frequency response models encoded in the UC that differentiate existing FCUC methods. Therefore, it is important to compare these models. As we discussed in the introduction, the SFR models in existing FCUC mainly focus on approximation of the system frequency or so-called center-of-inertia (COI) frequency. Currently, there are mainly two types of models to approximate the COI frequency nadir: (1) swing equation with a piecewise linear mechanical power approximation \cite{Lee2013,Chavez2014,Teng2015,Prakash2018,Wen2016,Badesa2019,Chu2020}, (2) swing-turbine-governor model \cite{Ahmadi2014,Gholami2018,Muzhikyan2018,Nguyen2019,Zhang2020b}. The first type considers the turbine governor dynamics as an open-loop piecewise linear function in time. It can result in considerable approximation error as it omits the interaction between mechanical power and frequency. A typical approximation can be found below \cite{Chavez2014}
\begin{subequations}
	\begin{align}
	\Delta P_{m}(t) =
	\begin{cases}
	0 & \text{if $t < t_{DB}$}\\
	\frac{R_{g}}{T_{d}}(t-t_{DB}) & \text{if  $T_{d} + t_{DB} \geq t \geq t_{DB}$}\\
	R_{g} & \text{$t \geq T_{d} + t_{DB}$}
	\end{cases}       
	\end{align}
\end{subequations}
where $t_{DB}$ represents the time when frequency deviation reaches the dead-band, $T_{d}$ is the delivery time of frequency response, and $R_{g}$ is the primary frequency response provision. And the system frequency is derived using the swing equation
\begin{subequations}
	\begin{align}
	& \frac{\text{d}\Delta f}{\text{d}t} = \frac{1}{2H_{\text{coi}}} (\Delta P_{m}- \Delta P_{d})
	\end{align}
\end{subequations}
where $\Delta P_{d}$ denotes the disturbance and $H_{\text{coi}}$ is the COI inertia constant. The COI inertia constant $H_{\text{coi}}$ is calculated as
\begin{align}
\begin{aligned}
H_{\text{coi}}=\dfrac{\sum_{i\in \mathcal{S}}^{N_{s}}(S_{i}^{s}H_{i}^{s})}{\sum_{i\in \mathcal{S}}^{N_{s}}S_{i}^{s}}
\end{aligned}
\end{align}
where $S_{i}^{s}$ and $H_{i}^{s}$ are the base and inertia constant of synchronous generator $i$, respectively. 
The second type uses the classic SFR model to capture the COI frequency. The SFR model usually connects the swing equation with an aggregated turbine-governor response model. A typical model can be found below
\begin{subequations}
	\begin{align}
	& \frac{\text{d}\Delta f}{\text{d}t} = \frac{1}{2H_{\text{coi}}} (\Delta P_{m}- \Delta P_{d})\\
	& \frac{\text{d}\Delta P_{m}}{\text{d}t} = \frac{1}{T_{3}} [-\Delta P_{m} + \Delta P_{v} + \frac{T_2}{T_1}(-\Delta P_{v}-\frac{1}{R}\Delta f)]\\
	& \frac{\text{d}\Delta P_{v}}{\text{d}t} = \frac{1}{T_{1}}(-\Delta P_{v} -\frac{1}{R}\Delta f)
	\end{align}
\end{subequations}
We compare the frequency dynamics that are captured by the aforementioned model-based approximation methods and the DAE model, which our data-driven approach is able to encode into the FCUC.  To do this, we simulate the system under a generator outage disturbance and plug the electric power deviation obtained from the DAE simulation into the two approximation models. The mechanical power and frequency dynamics are compared in Fig. \ref{fig_dyn_comp_pm} and Fig. \ref{fig_dyn_comp_f}, respectively. As shown, the first method admits significant error since it does not capture the full dynamic characteristics of the mechanical power. The second method outperforms the first one but still has around 0.1 Hz error. Besides the approximation error, other limitations of these encoding models have been discussed in the introduction.

\subsection{On Larger Test Systems}\label{sec_discuss_sub_sys}
Efforts have been made to apply the FCUC on large-scale test systems, including NPCC 140-bus system \cite{wenyun}\cite{Cui2021} and WECC 179-bus system \cite{wangbin_2016} \footnote{The open-source system data can be found \url{https://github.com/cuihantao/andes/tree/master/andes/cases}.}. However, in such systems, the generators are dense aggregations of original small generators, and in most cases, the system cannot withstand the worst-case generation outage if there are up to three de-committed generators. The labeling results of the NPCC system for 5000 samples are shown in Table \ref{tab_3.3}. As we can see, the stability condition becomes the binding constraint rather than the frequency security, which is because of the generator aggregations. FCUC, or more generally dynamic-constrained operation, on such systems, can result in overly conservative results. Open-source large-scale test systems that are suitable for studying FCUC and the interaction between dynamics and operation under high renewable penetration are needed by the research community.
\begin{table}[H]
	\caption{Labeling Results of the NPCC System}
	\centering
	\begin{tabular}{lclclclclclclcl}
		\toprule 
		Total & Unsolvable & Unstable  & Stable & Averaged Freq. Nadir \\
		\midrule
		5000 & 2293 & 2647 & 60 & -0.4988 (Hz) \\
		\bottomrule
	\end{tabular}
	\label{tab_3.3}
\end{table}

\subsection{On Prediction Error}\label{sec_discuss_sub_error}
It is worth noting the prediction error with negative residual error ($f_{\text{ndr}}-\hat{f}_{\text{ndr}}<0$, where $f_{\text{ndr}}<0$ and $\hat{f}_{\text{ndr}}<0$ are the actual and predicted frequency deviation nadirs from the nominal value) could possibly compromise the frequency security and trigger UFLS. One immediate remedy is to collect the empirical prediction errors and add an extra security margin in the FCUC in Eq. (\ref{eq_enforce_fc_1}). As for the stability prediction, our observation is that predicting post-fault stability only using the steady-state operation information can be challenging. Most of the machine learning-based dynamic security/stability prediction methods will require on-fault trajectories and the beginning of the post-fault trajectories. Due to this prediction challenge, strictly enforcing Eq. (\ref{eq_fcuc_stab_4}) may result in overly conservative results, and relaxation as conducted in Eq. (\ref{eq_enforce_fc_2}) is needed.

\subsection{On Cost Benefit}\label{sec_discuss_sub_cost}
Based on Table \ref{tab_nadir}, the post-contingency frequency responses are just above the UFLS threshold, indicating that the \emph{minimum adequate generators} are committed and a good trade-off between cost and security. The FCUC will result in a higher day-ahead UC cost when wind power in the system is abundant but may lead to a lower intra-day operation cost, especially for secondary and tertiary frequency control, depending on the uncertainty realization of the renewable sources. Traditional UC may rely on fast responsive units to compensate for uncertainty, which is expensive and have limited capacity. With an insufficient number of responsive units, generator outages can result in catastrophic outcomes and lead to more costs in customer interruption and restoration \cite{NG_frequency}.

\subsection{On Data Bias}\label{sec_discuss_sub_bias}
The data bias issue can compromise the generalization of the prediction models. Two typical data biases that we possibly face are sample bias and recall bias. Sample bias occurs when a dataset does not reflect the realities of the environment in which a model will run. Recall bias arises when you label similar types of data inconsistently. We can resolve the sample bias issue by sampling from wide-range operation conditions. To traverse as many conditions as possible, we sample generator power from their dispatch intervals in the uniform distribution and load consumption in the normal distribution with a large variance, as discussed in Section \ref{sec_al_sampling}. As for the recall bias, a post-process is proposed to filter out numerical transients in bus frequencies from physical ones such that incorrect label values are prevented, as discussed in Section \ref{sec_sub_process}.

\subsection{On Computation Performance}\label{sec_discuss_sub_computation}
The high-fidelity power system simulation is conducted using PSS/E on a window laptop with an Intel Core i7 1.9 GHz CPU and a 16 GB RAM. The neural network training and FCUC are performed using Tensorflow 1.14 and Gurobi 9.1, respectively, on a MacBook Pro with an Intel Core i9 2.3 GHz CPU and a 16 GB RAM. The averaged computation time is given as follows
\begin{itemize}
	\item 2.53[s] in labeling each sample
	\item 188.61[s] in a 3000-epoch training for each predictor
	\item 269.49[s] in FCUC
\end{itemize}
Considering the fact that this effort is to replace part of the iterative process between traditional UC and DSA, the computation performance is satisfactory.

\section{Conclusions}\label{sec_con}
The UC formulation without frequency response constraints will lead to insecure and unstable dynamic responses under contingencies when the renewable penetration becomes high. The existing FCUC methods are either incapable of considering realistic models and corresponding stability conditions. The paper presents a generic deep learning-based framework for FCUC. Two neural network models are constructed to represent the frequency nadir and stability characteristics given a specific operating condition and a worst-case contingency. The proposed region-of-interest active sampling effectively selects power injection samples whose frequency nadirs are closer to the UFLS threshold. Such a dataset later proves to be beneficial for training accuracy improvement. The FCUC formulation can effectively schedule frequency secure generator commitments, which is verified using the PSS/E simulation.

\appendices
%\section{Parameters}\label{appendix_para}
%Diesel generator: Rated power: 2 [MW], $H_{D,i}=1$ [s], $\tau_{d,i}=0.2$ [s], $\tau_{sm,i}=0.1$ [s] for $i=1,2$.
%Wind turbine generator: Rated power: 1 [MW], $H_{T,i}=4$ [s], $K_{P,i}^{T}=2$, $K_{I,i}^{T}=0.1$, $K_{P,i}^{Q}=1$, $K_{I,i}^{Q}=5$, $K_{P,i}^{C}=0.6$, $K_{I,i}^{C}=8$ for $i=1,2,3$.
%\section{Data of System Frequency Response Model}\label{appendix_SFR}

\section{Linearized AC Power Flow}\label{appendix_pl}
Consider the original AC power flow equation as follows
\begin{subequations}
	\begin{align}
	&\begin{aligned}
	&P_{ij,t}=G_{ij}(V_{i,t})^{2} - G_{ij}[V_{i,t}V_{j,t}\cos(\theta_{ij,t}))] \\
	&\quad\quad\quad\quad- B_{ij}[V_{i,t}V_{j,t}\cos(\theta_{ij,t})]\quad \forall i,j,i\neq j,\forall t
	\end{aligned}\\
	&\begin{aligned}
	&Q_{ij,t}=-B_{ij}(V_{i,t})^{2} + B_{ij}[V_{i,t}V_{j,t}\cos(\theta_{ij,t})] \\
	&\quad\quad\quad\quad- G_{ij}[V_{i,t}V_{j,t}\sin(\theta_{ij,t})] \quad \forall i,j,i\neq j,\forall t\\
	\end{aligned}
	\end{align}
\end{subequations}
Note that $G_{ij}$ and $B_{ij}$ are the line conductance and susceptance, respectively, instead of the entry of the $Y$ matrix. The following linearization terms in \cite{trodden2014optimization} are considered
\begin{subequations}
	\begin{align}
	& V_{i,t}^{2} \approx 2V_{i,t}-1\\
	& V_{i,t}V_{j,t}\cos(\theta_{ij,t}) \approx V_{i,t} + V_{j,t} - 1\\
	& V_{i,t}V_{j,t}\sin(\theta_{ij,t}) \approx \theta_{ij,t}=\theta_{i,t} - \theta_{j,t}
	\end{align}
\end{subequations}
Substituting these terms into the AC power flow equations yields
\begin{subequations}
	\begin{align}
	&\begin{aligned}
	&P_{ij,t}=G_{ij}(2V_{i,t}-1) - G_{ij}(V_{i,t} + V_{j,t} - 1) \\
	&\quad\quad\quad\quad- B_{ij}(\theta_{i,t} - \theta_{j,t})\quad \forall i,j,i\neq j,\forall t
	\end{aligned}\\
	&\begin{aligned}
	&Q_{ij,t}=-B_{ij}(2V_{i,t}-1) + B_{ij}(V_{i,t} + V_{j,t} - 1) \\
	&\quad\quad\quad\quad- G_{ij}(\theta_{i,t} - \theta_{j,t}) \quad \forall i,j,i\neq j,\forall t\\
	\end{aligned}
	\end{align}
\end{subequations}

\bibliography{IEEEabrv_zyc, library, Ref_AL_UC}
\bibliographystyle{IEEEtran}

\end{document}